\documentclass[
prx,twocolumn,superscriptaddress, amssymb, amsmath]{revtex4-1}
\usepackage{xcolor}
\usepackage{graphicx}
\usepackage{dcolumn}
\usepackage{bm}
\usepackage{amsmath}
\usepackage{lipsum}
\usepackage{mathtools}
\usepackage{tikz}
\usepackage{soul}
\usepackage[normalem]{ulem}
\usepackage{dsfont}
\usepackage{braket}
\usepackage[colorlinks]{hyperref}

\newcommand{\be}{\begin{equation}}
\newcommand{\ee}{\end{equation}}

\DeclareMathOperator{\Tr}{Tr}

\newcommand{\beq}{\begin{equation}}
\newcommand{\eeq}{\end{equation}}
\newcommand{\beqn}{\begin{eqnarray}}
\newcommand{\eeqn}{\end{eqnarray}}

\usepackage[noframe]{showframe}
\usepackage{framed}

\renewenvironment{shaded}{%
  \MakeFramed{\advance\hsize-\width \FrameRestore\FrameRestore}}%
 {\endMakeFramed}
\definecolor{shadecolor}{gray}{0.75}

\begin{document}
\newcommand{\jianhao}[1]{ { \color{violet} \small (\textsf{JHZ}) \textsf{\textsl{#1}} }}
\newcommand{\zhen}[1]{ { \color{blue} \small (\textsf{ZB}) \textsf{\textsl{#1}} }}
\newcommand{\yy}[1]{ { \color{red} \small (\textsf{YY}) \textsf{\textsl{#1}} }}
\newcommand{\fp}[1]{ { \color{green} \small (\textsf{FP}) \textsf{\textsl{#1}} }}

\newcommand{\pab}[1]{ { \color{olive} \small (\textsf{PS}) \textsf{\textsl{#1}} }}
\newcommand{\olive}[1]{ {\color{olive}#1}}

\title{Entanglement Holography in Quantum Phases via Twisted Rényi-N Correlators}

\author{Pablo Sala}

\affiliation{Department of Physics and Institute for Quantum Information and Matter, California Institute of Technology, Pasadena, California 91125, USA}

\affiliation{Walter Burke Institute for Theoretical Physics, California Institute of Technology, Pasadena, CA 91125, USA}

\author{Frank Pollmann}

\affiliation{Technical University of Munich, TUM School of Natural Sciences, Physics Department, 85748 Garching, Germany}

\affiliation{Munich Center for Quantum Science and Technology (MCQST), Schellingstr 4, 80799 Munchen, Germany}

\author{Masaki Oshikawa}

\affiliation{Institute for Solid State Physics, University of Tokyo, Kashiwa, Chiba 277-8581, Japan }

\author{Yizhi You}

\affiliation{Department of Physics, Northeastern University, Boston, MA, 02115, USA}

\date{\today}
\begin{abstract}
 We introduce a holographic framework for the entanglement Hamiltonian in symmetry-protected topological (SPT) phases with area-law entanglement, whose reduced density matrix \(\rho = e^{-H_e}\) can be treated as a lower-dimensional mixed state. By replicating \(\rho\), we reconstruct the fixed-point SPT wavefunction, establishing an exact correspondence between the bulk strange correlator of the \((d+1)\)-dimensional SPT state and the twisted Rényi-\(N\) operator of the \(d\)-dimensional reduced density matrix. Notably, the reduced density matrix exhibits long-range or quasi-long-range order along the replica direction, revealing a universal entanglement feature in SPT phases. As a colloary, we generalized the framework of twisted Rényi-\(N\) correlator to thermal states and open quantum systems, providing an alternative formulation of the Lieb-Schultz-Mattis theorem, applicable to both closed and open systems.   
Finally, we extend our protocol to mixed-state SPT phases and introduce new quantum information metrics—twisted Rényi-N correlators of the surgery operator—to characterize the topology of mixed states.
\end{abstract}

\maketitle
\section{Introduction}
Entanglement underpins quantum many-body theory, offering a powerful quantum information lens for understanding complex many-body phenomena~\cite{RevModPhys.80.517,calabrese2008entanglement,calabrese2009entanglement,Peschel_2009,pollmann2012symmetry,Verstraete9,Schollwoeck11,PhysRevB.84.195103,PhysRevLett.105.115501,PhysRevLett.104.130502,chandran2014universal,PhysRevLett.110.236801,zhang2012quasiparticle,KitaevPreskill,Levin-2006,luitz2014participation}.  Extensive investigations have shown that key topological characteristics—including quasiparticle statistics, edge excitations~\cite{chen2010local,chen2012symmetry,Chen2011-kz,Chen2011-et,Pollmann2010,Pollmann2012-lv,pollmann2012symmetry,Turner2011-zi,Li-2008,Peschel_2009,pollmann2012symmetry,PhysRevB.84.195103,wybo2021visualizing,PhysRevLett.105.115501,PhysRevLett.104.130502,chandran2014universal,PhysRevLett.110.236801,nakagawa2017capacity,you2022observing}, central charge, and topological Berry phase—are encoded in the entanglement spectrum (ES)~\cite{Li-2008,zhang2012quasiparticle,tu2013momentum,matsuura2016charged,marvian2017symmetry,you2020higher}. Central to these studies is the reduced density matrix (RDM) $\rho_A$, obtained by tracing out part of the system to focus on the entanglement of a chosen subsystem. The eigenvalues of $\rho_A$ make up the entanglement spectrum (ES), and the entanglement Hamiltonian $H_E = -\ln \rho_A$ encodes entanglement energy levels that are closely linked to boundary excitations.

While extensive attention has centered on the spectrum of the entanglement Hamiltonian~\cite{luitz2014participation,santos2018symmetry,lundgren2014momentum,sohal2020entanglement,henderson2021entanglement,nakagawa2017capacity,nehra2020flat,tu2013momentum,mollabashi2014entanglement,nehra2020flat}, a pressing question is whether one can uncover the physical observables of the bulk wavefunction such as strange correlators~\cite{you2014wave,vanhove2018mapping,wu2015quantum,vanhove2018mapping,zhou2022detecting,lepori2023strange} from it, e.g., by considering non-linear functions of the RDM. In systems that obey an area‑law, the RDM acts as a lower‑dimensional mixed‑state ensemble; evaluating quantum‑information metrics for this ensemble uncovers hidden features of the original wavefunction. From an alternative perspective, the RDM of a symmetry protected topological (SPT) wavefunction with symmetry \( G \) can be understood through a `cutting and gluing' approach introduced in Ref.~\cite{qi2012general,chen2013quantum,lundgren2013entanglement}. In this approach, the two parts of a bipartite system are coupled, as illustrated in Fig.~\ref{cutandglue}. Although each edge state is individually anomalous under the symmetry \( G \) (resulting in a gapless or degenerate energy spectrum), coupling the two edges opens a gap, causing the boundaries to merge and become part of the bulk wavefunction.
By treating the one boundary as ancilla qubits (part of the environment) and the other boundary (along with the connected bulk) as the system of interest, the coupling between the edges can be viewed as a quantum channel in an open system. This channel decoheres the anomalous gapless edge at the bottom, converting it into a mixed state. Notably, the resulting mixed-state density matrix exactly matches the RDM \( \rho \) of the SPT ordered wavefunction under a spatial cut. 
This perspective offers a framework to investigate the effects of decoherence and dissipation on an anomalous gapless boundary by analyzing the RDM of an SPT wavefunction\cite{murciano2023measurement,garratt2024probing,tang2024critical,yang2023entanglement,sun2024holographic,sala2024spontaneous}. 

Spurred by rapid developments in open quantum systems~\cite{zhou2025reviving,huang2025interaction,ding2024boundary,hsin2024anomalies,sala2024spontaneous} driven by decoherence and dissipation, several nonlinear observables—such as coherent information and Rényi-\(2\) correlations—have been introduced to characterize mixed-state ensembles~\cite{bao2023mixed,fan2023diagnostics,lee2022symmetry,lee2024exact,zhang2022strange,lessa2024mixed}. A natural question arises: Can these nonlinear measures be extended to the RDM of SPT wavefunctions? If so, do they capture the essential bulk properties of these systems?
Recent work \cite{zang2024detecting,wang2024anomaly} has partially addressed this question by examining a class of mixed states that exhibit anomalies under a weak symmetry \(G\) and proposes characteristic signatures to diagnose such anomalies in mixed-state ensembles. Remarkably, the anomalous action of \(G\) on these mixed-state ensembles closely resembles the symmetry action on the RDM of a \(G\)-symmetric SPT wavefunction. This resemblance suggests that such anomalous mixed states can arise as the RDM of a pure state in a higher-dimensional system.

In this work, we develop a holographic framework for the entanglement Hamiltonian in topological phases both in and out of equilibrium. For a \((d+1)\)-dimensional wavefunction exhibiting area-law entanglement, its reduced density matrix under spatial bipartition yields a \(d\)-dimensional mixed state. This motivates us to build a holographic duality that reconstructs the \((d+1)\)-dimensional SPT wavefunction by replicating its \(d\)-dimensional RDM, \(\rho = e^{-H_e}\), along the replica direction. This construction establishes an exact correspondence between two seemingly unrelated objects: the bulk strange correlator of the \(d+1\)-dimensional SPT wavefunction and the twisted Rényi-\(N\) correlator (TRNC) of the RDM. Remarkably, we find that the RDM of an SPT state supports long-range or quasi-long-range order along the `replica direction', precisely captured by the TRNC. Our approach introduces a new quantum information metric for investigating SPT wavefunctions by examining the nonlinear properties of the entanglement spectrum.

The rest of this paper is organized as follows.  
In Section~\ref{sec:Sec_II}, we review the holographic framework introduced in Ref.~\cite{sun2024holographic}, which constructs a $(d+1)$-dimensional fixed‑point SPT state by replicating its $d$-dimensional RDMs. This framework establishes a connection between twisted Rényi-\(N\) correlators of the RDM and bulk strange correlations. In Section~\ref{sec:sec_III}, we numerically compute the twisted Rényi-\(N\) correlators in 1d SPT phases, demonstrating their ability to distinguish different SPT states and detect phase transitions.  
In Section~\ref{sec:sec_IV}, we extend our analysis to 2d systems, including Chern insulators and quantum spin Hall type states, using both analytical arguments and numerical simulations. We also extend the framework of the reduced density matrix and its twisted Rényi-\(N\) correlator to thermal states and open quantum systems, using it as a corollary or alternative proof of the Lieb–Schultz–Mattis theorem for both closed and open systems in Sec.~\ref{sec:corollary}.
In Section~\ref{sec:sec_V}, we extend our formalism to open quantum systems, demonstrating that the topology of mixed-state symmetry-protected topological (mSPT) phases can be characterized using twisted Rényi-N correlators of the lower-dimensional surgery operator.

\section{Mapping between SPT wavefunction  and its RDM} \label{sec:Sec_II}

\subsection{Notion of symmetries for the density matrix}
We say that a quantum state is symmetric under a symmetry group $\mathcal{G}$, if for every element $g\in  \mathcal{G}$, the action of the group $U:g\in \mathcal{G}\to \mathcal{L}(\mathcal{H})$ on the Hilbert space $\mathcal{H}$ leaves the state invariant, namely
\begin{align} \label{eq:symm_psi}
   U_g |\Psi \rangle= e^{i\theta}|\Psi \rangle
\end{align}
With $e^{i\theta}$ a global phase.
When considering a mixed state characterized by a density matrix, a symmetry can be manifested in two distinct ways, denoted as strong (or exact) and weak (or average) symmetry. A mixed state is weakly-symmetric under $\mathcal{G}$, if the density matrix is invariant under the symmetry transformation \( U_g \), acting from both the left (ket) and the right (bra):
\begin{align} \label{eq:weak}
    U_g \rho U^\dag_g = \rho.
\end{align}
Such a symmetry transformation can be interpreted as implementing the symmetry operation to both the ket and bra space of the density matrix. For example, this is the case for the RDM $\rho$ of a symmetric many-body wavefunction $|\Psi \rangle$ that is invariant under a symmetry transformation \(U_g\). When this is the case, we can write $\rho$ as
$\rho = \sum_\nu \lambda_\nu | \nu \rangle \langle \nu |$, where $|\nu \rangle$
represents different eigenstates of the symmetry transformation $U_g$.


 The second notion of a symmetry for a mixed state is denoted as strong symmetry. It requires the density matrix to be invariant under the symmetry transformation \( U_g \), acting either from the left or from the right:
\begin{align}
     U_g \rho=e^{i\theta}\rho, \,\,\,\, \rho U_g^\dagger=e^{-i\theta}\rho
\end{align}
with $e^{i\theta}$ being a global phase. Notice that if a state is strongly symmetric, then it is also weakly symmetric. Canonical examples of symmetric density matrices are given by the grand canonical and canonical ensembles, which are weakly and strongly symmetric, respectively.

\subsection{Recover the full wavefunction from the reduced density matrix}\label{sec:recover}

\begin{figure}[h]
\centering
\begin{tikzpicture}
\node at (0,0){\includegraphics[scale=0.9]{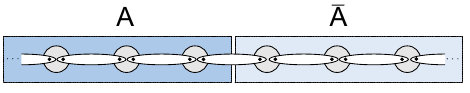}};
\node at (0.25,-3){\includegraphics[scale=0.35]{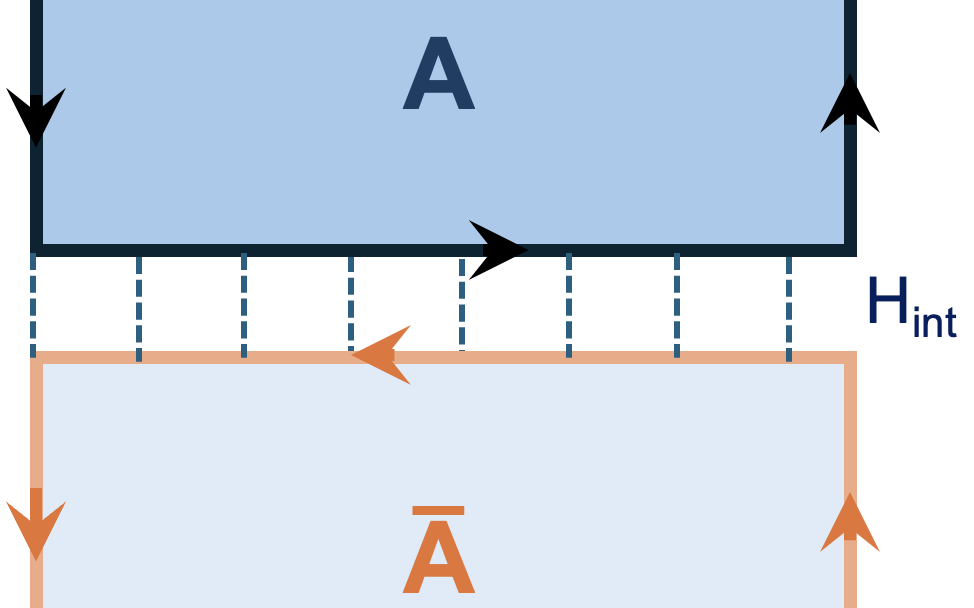}};
\node at (-3.8,0) {(a)};
\node at (-3.8,-1.3) {(b)};
\end{tikzpicture}
    \caption{a) Spatial bipartition of a 1d SPT into regions \( A \) and \( \bar{A} \). The reduced density matrix \( \rho \) resembles a 0d mixed state with degrees of freedom localized near the cut.
    (b) The cut-and-glue process begins by splitting the Chern insulator into two disconnected regions, \( A \)(top) and \( \bar{A} \)(bottom), each hosting chiral edge states. Interactions are then reintroduced, merging the edges into the bulk, analogous to coupling chiral propagating bosons on a ladder.} 
    \label{cutandglue}
\end{figure}

The RDM of an SPT wavefunction is obtained by tracing out partial degrees of freedom, resulting in a lower-dimensional mixed state. Here, we offer an alternative viewpoint: the RDM can be understood as a mixed-state ensemble arising from decohering the system’s gapless boundary modes.
Concretely, consider a \((d+1)\)-dimensional SPT wavefunction with open boundaries that host anomalous, gapless edge modes (see Fig.~\ref{cutandglue}b). When these two boundaries are reconnected by adding inter-edge interactions (effectively merging them into the bulk), the top edge can be viewed as an ancilla, and the bottom edge, together with the bulk, becomes the system of interest. The inter-edge interactions act like a decohering quantum channel, transforming the wavefunction at the bottom into a \(d\)-dimensional mixed-state ensemble. This mixed state is precisely the RDM \(\rho\) associated with spatially partitioning the top and bottom regions.

In this interpretation, the RDM naturally emerges as the mixed state ensemble of a noisy quantum channel acting on the bottom boundary qubits, with Kraus operators determined by the inter-edge coupling. This perspective highlights how tracing out part of the system can be viewed as physically decohering the boundary modes, yielding a mixed-state ensemble in lower dimensions. These observations raise several key questions:
\begin{enumerate}
    \item Can the salient properties of the bulk wavefunction be inferred from measurements of its reduced density matrix?
    
    \item Is it possible to map nonlinear observables of the reduced density matrix to corresponding bulk correlation functions?

    \item When the boundary of an SPT state undergoes decoherence and becomes a mixed state, does it retain its distinctive topological or symmetry-protected features?
\end{enumerate}

Motivated by these questions, we show that a \(d\)-dimensional density matrix \(\rho\) with an anomaly under a weak symmetry \(G\) can be extended to a \((d+1)\)-dimensional SPT wavefunction \(\Psi^{\mathrm{spt}}\) protected by \(G\) symmetry. This dimensional-extension mapping, originally proposed in Ref.~\cite{sun2024holographic}, transforms a density matrix into a higher-dimensional wavefunction through replication.

\begin{figure}[h]
    \centering
\begin{tikzpicture}
\node at (0,0){\includegraphics[scale=0.5]{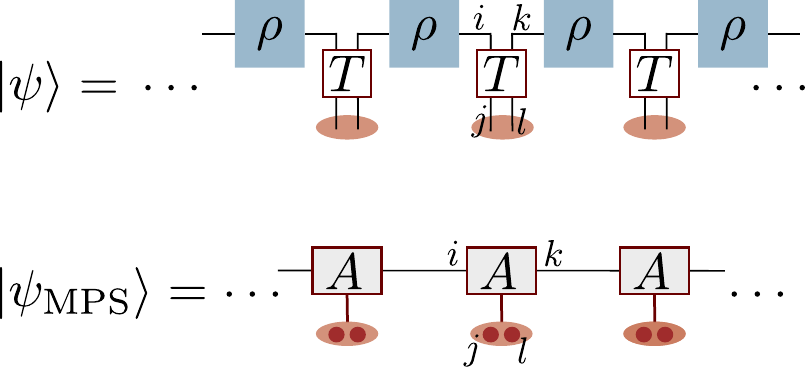}};
\node at (-3.8,1.5) {(a)};
\node at (-3.8,-0.5) {(b)};
\end{tikzpicture}
    \caption{a) The 1d SPT wavefunction is constructed by taking replica copies of the 0d density matrix with
     \( \rho = \sum_\nu \lambda_\nu |\nu \rangle \langle \nu | \)
    and inserting a rank-4 \(T\)-tensor between each pair. In this construction, the bra from the \(i\)th copy and the ket from the \((i+1)\)th copy form a single unit cell (red oval), naturally yielding an MPS description of the wavefunction.  
b) The corresponding MPS tensors can be written as \(A^{jl}_{ik} \sim \delta_{ij}\,\delta_{kl}\,\lambda_k \,\lambda_i\).}
    \label{wavefunc}
\end{figure} 

The dimensional-extension mapping proceeds by replicating a vectorized \( d \)-dim density matrix
\( \rho = \sum_\nu \lambda_\nu | \nu \rangle \langle \nu | \),
where \( | \nu \rangle \) is a basis state of the \( d \)-dimensional system, and $\lambda_\nu $ are non-negative real coefficients. Since we will later replicate the density matrix into \( N \) copies, each replica of the density matrix is labeled as \( \rho_i \) with an index \( i = 1, \dots, N \). The replicas are then glued together along an additional dimension (the `replica dimension') to form a pure state in \( d+1 \) dimensions. This requires two important steps that are reminiscent of the construction of the Affleck-Kennedy-Lieb-Tasaki (AKLT)\cite{Affleck1987-wn} wave function: 

\begin{enumerate}
    \item \textbf{Build up entanglement between sites.} We consider a density matrix $\rho_i$ and map it to its Choi-double state (a procedure known as vectorization), defined as  
$|\rho\rangle\rangle_i = \sum_\nu \lambda_\nu |\nu \rangle_{L,i} | \nu \rangle_{R,i} $
which resides in a Hilbert space with doubled dimension. The resulting entangled state provides the Schmidt eigenvalues in spatial partitions.

    \item \textbf{Fix the local Hilbert space.} Insert a rank-4 tensor between the replicas as in Fig.~\ref{wavefunc}, which defines the local Hilbert space. This could be simply $\mathds{1}$ or some projector, as in the case of the spin-$1$ AKLT wave function. 
The former requires to fix a basis, which we assume to be the $z$-local computational basis.
\end{enumerate}

To proceed, we begin with a specific example for \(d=0\), while noting that the approach readily generalizes to higher dimensions. 
In this construction, the bra vector of the \(i\)-th \(\rho\), i.e., $|\nu \rangle_{R,i}$ and the ket vector of the \((i+1)\)-th \(\rho\) given by $|\nu \rangle_{L,i+1}$ together define a \textit{unit cell} (labeled by the index $x_i=i+\frac{1}{2}$). We can then insert a rank-$4$ tensor \(T\) to introduce an additional intra-unit-cell coupling. For simplicity, we choose \(T_{ijkl} = \delta_{ij}\,\delta_{kl}\) in the following, yielding the fixed-point state \(|\Psi^{\text{SPT}}\rangle\). Likewise, the bra and ket spaces of each \(\rho\) form a `entangling block' that entangles the right component of the \(i\)-th site with the left component of the \((i+1)\)-th site. 
 This setup effectively extends the 0$d$ mixed state into a 1$d$ wave function, where the replica index plays the role of the spatial site index, and is in fact reminiscent of the matrix product state construction of the AKLT wavefunction.
Following this procedure, the wave function is expressed as a tensor product of entangled pairs between sites as:
\begin{align}\label{fixed}
&|\Psi^{\text{SPT}}\rangle= \sum_{\{\nu_i\}} \prod_{i=1}^N \Lambda_{\nu_i} \bigotimes_{i=1}^N   \underbrace{(|\nu_i\rangle_{R,i} |\nu_{i+1}\rangle_{L,i+1} )}_{\text{unit-cell } x_i=i+\frac{1}{2}},
\end{align}
Where $\Lambda_{\nu}$ are defined as $\Lambda_{\nu}\equiv \lambda_{\nu}/\sqrt{\sum_\nu {\lambda_\nu}^2}$ such that $|\Psi^{\text{SPT}}\rangle$ is normalized. As a result, the many-body state \(|\Psi^{\text{SPT}}\rangle\) corresponds to a product of entangled pairs across all sites. The wavefunction \(|\Psi^{\text{SPT}}\rangle\) naturally admits a matrix product state (MPS) representation illustrated in Fig.~\ref{wavefunc}. The corresponding MPS tensors can be expressed as \(A^{jl}_{ik} \sim \delta_{ij} \, \delta_{kl} \, \lambda_k \, \lambda_i\). 
By construction, the pure wave function in Eq.~\eqref{fixed} inherits the weak \(G\) symmetry of the mixed state \(\rho\), which acts as:
\begin{align}\label{eq:strongsymact}
U^{(N)}_h|\Psi^{\text{SPT}}\rangle= \sum_{\{\nu_i\}} \prod_{i=1}^N\Lambda_{\nu_i} \bigotimes_{i=1}^N (U^*_{h} U_h)_i  (|\nu_i\rangle_{R,i} |\nu_{i+1}\rangle_{L,i+1} )
\end{align}
for every $h\in G$. Notice that the symmetry operators \(U_h^*\) act on the R component in each unit cell in a conjugate manner as they originate from the bra of the density matrix.  Since the fixed-point wave function in Eq.~(\ref{fixed}) is a tensor product of entangled pairs, it is straightforward to see that the resulting state \(|\Psi^{\text{SPT}}\rangle\) possesses a global symmetry \(G\). 

Moreover, through this mapping, we can also easily compute the bipartite entanglement entropy. In particular, for a cut between unit cells $x_{i-1}=i-\frac{1}{2}$ and $x_{i}=i+\frac{1}{2}$, the entanglement entropy is given by
$-\sum_{\nu_i} \Lambda^2_{\nu_i} \ln(\Lambda^2_{\nu_i})$. 
Furthermore, this dimensional extension--from a lower-dimensional reduced density matrix to a higher-dimensional wavefunction--implies that the observable of the wavefunction can be fully captured by analyzing non-linear quantities within the lower-dimensional reduced density matrix.

\subsection{Twisted Rényi-N correlator: Long range order in the replica direction}\label{sec:twisted}

To leverage this mapping, we demonstrate that the salient properties of a \((d+1)\)-dim SPT wavefunction are dual to the nonlinear observables derived from the d-dim reduced density matrix. To clarify this relationship, we revisit the concept of the \textit{twisted Rényi-\(N\) correlator} of the reduced density matrix, initially proposed in Ref.~\cite{sun2024holographic}. In Ref.~\cite{sun2024holographic}, the \textit{twisted Rényi-\(N\) correlator} was introduced as a quantum-information quantity to probe mixed states in open quantum systems subject to dissipation and decoherence. In this work, we demonstrate that this correlator has broader applicability and can also be employed to analyze reduced density matrices in SPT wavefunctions.

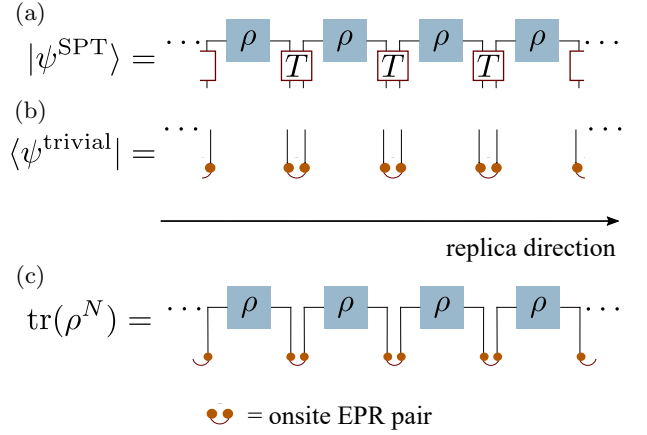
\begin{figure}[h]
    \centering
\begin{tikzpicture}
\node at (0,0){\includegraphics[scale=0.49]{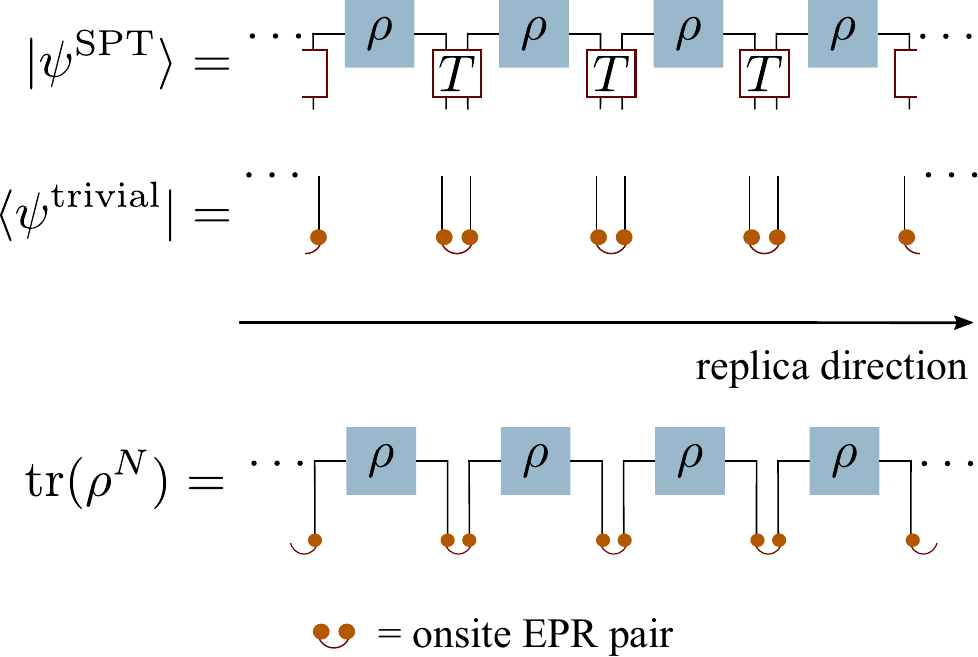}};
\node at (-3.8,2.8) {(a)};
\node at (-3.8,1.5) {(b)};
\node at (-3.8,-0.7) {(c)};
\end{tikzpicture}
    \caption{The inner product between the SPT wavefunction and a trivial on-site entangled state, illustrated in panel a), is mathematically equivalent to the trace of \(\rho^{2N}\), as depicted in panel b.}
    \label{sc1}
\end{figure}

To establish the framework, we begin by observing that the inner product between the fixed-point wavefunction $|\Psi^{\text{SPT}}\rangle$ under periodic boundary conditions in Eq.~(\ref{fixed}) and a trivial tensor product state (with no entanglement between unit cells) can be compactly expressed in terms of a trace. In particular, let us label a unit cell by $x$ 
(labeled by $i+\frac{1}{2}$ in the previous section) with $x=1,\dots, N$ assuming periodic boundary conditions, and consider the trivial state (see Fig.~\ref{sc1}a):
 \begin{align}
|\Psi^{\text{trivial}}_M\rangle = \bigotimes_{x}  \left(\sum_{a,b} M_{ab} |a\rangle_{L,x} |b\rangle_{R,x} \right).
\end{align}
Then, the inner product between these two wave functions is given by the trace
 \begin{align}
&\langle \Psi^{\text{trivial}}_M  |\Psi^{\text{SPT}}\rangle = \text{Tr}[\rho M \rho  M  \rho M\rho M...\rho M\rho M],
\end{align}
which depends on the choice of the trivial state via \(M\).
A common example is to take \(M\) as the identity matrix, for which \(|\Psi^{\mathrm{trivial}}\rangle\) can be written as a tensor product of on-site EPR states: 
\[
\bigl|\Psi^{\text{trivial}}\bigr\rangle 
= \mathcal{A}\bigotimes_x \left(\sum_g \bigl|g\bigr\rangle_{L,x}\, \bigl|g\bigr\rangle_{R,x} \right),
\]
with \(\mathcal{A}\) being the normalization constant. Notice that this trivial state is manifestly \(G\)-symmetric, and that other trivial symmetric states can be similarly used based on specific needs.

\begin{figure}[h]
    \centering
    \begin{tikzpicture}
\node at (0,0){\includegraphics[scale=0.42]{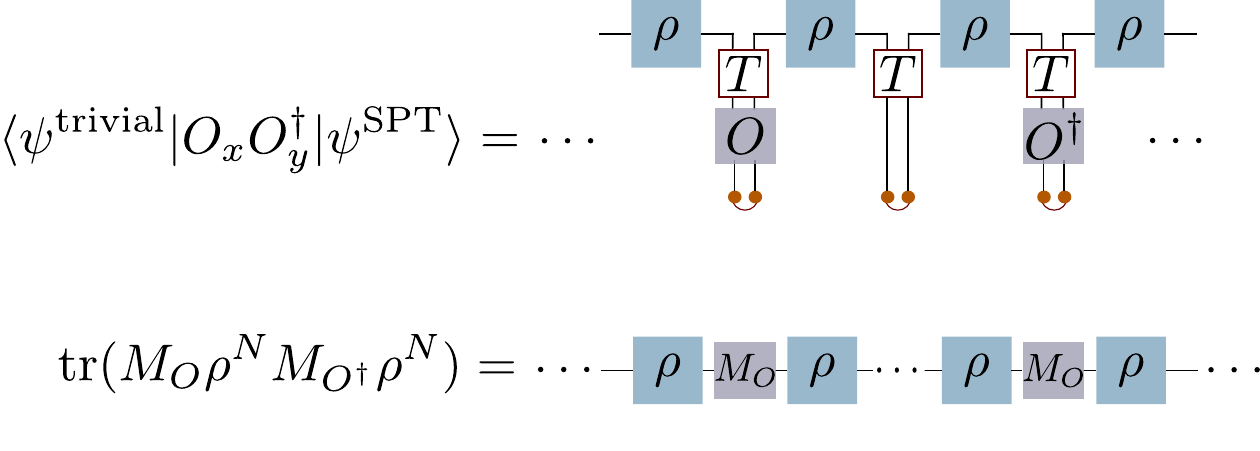}};
\node at (-3.5,1.5) {(a)};
\node at (-3.5,-0.5) {(b)};
\end{tikzpicture}
    \caption{The strange correlator of the dual wavefunction, shown in panel a, is equivalent to the twisted Rényi-\(N\) correlator depicted in panel b.}
    \label{sc2}
\end{figure}

The twisted Rényi-N correlator is then defined as:
\begin{align}\label{eq:twistedm}
&C(N)_M=\frac{\text{Tr}[\rho  M_{O} (\rho M)^{N-1}\rho M^{\dagger}_{O}(\rho M)^{N-1}]}{\text{Tr}[(\rho M)^{2N}]},
\end{align}
From now on, we choose \( M \) to be the identity operator for \( C(N) \), so that the trivial state \( \Psi^{\mathrm{trivial}} \) is a fixed tensor product of on-site EPR pairs. Under this choice, the twisted Rényi-\( N \) correlator becomes:
\begin{align}\label{eq:twisted}
&C(N)=\frac{\text{Tr}[\rho^N  M_{O} \rho^N M^{\dagger}_{O}]}{\text{Tr}[\rho^{2N}]} .
\end{align}
which equals~\cite{sun2024holographic}
\begin{align}\label{eq:strange_corr}
\frac{\langle \Psi^{\text{trivial}} | O^{\dagger}_1 O_{N} |\Psi^{\text{SPT}}\rangle }{\langle \Psi^{\text{trivial}}  |\Psi^{\text{SPT}}\rangle } 
\end{align}
as shown in Fig.~\ref{sc2}. The matrix \(M_O\) in Eq.~\eqref{eq:twisted}, is defined by the specific observable \(O\) being measured and must satisfy the relation
\[
M_O \;=\; O_{ijkl}\,\delta_{ij}.
\]

In summary, the twisted Rényi-\(N\) operator associated with the reduced density matrix RDM naturally reproduces the bulk strange correlator of the higher-dimensional SPT wavefunction. Specifically, it measures the correlation between two operators placed along the \(x\)-axis while taking an inner product with a trivial state. Within the dimension extension mapping, the \(x\)-direction acts as the replica direction, and the operators insertion \(M_O\) and \(M_O^\dagger\) are separated by \(N\) copies of the density matrix. This construction allows to probe operator correlations across a distance \(N\) along the replica dimension.

Given that the SPT wavefunction exhibits long-range (or quasi-long-range) order in its strange correlator, the twisted Rényi-\(N\) operator is expected to display long-range order along the replica direction. Importantly, \(M_O\) and \(M_O^\dagger\) act on the same \textit{spatial} degrees of freedom of the reduced density matrix \(\rho\) but are separated along the replica axis. As a result, the twisted Rényi-N correlator of the RDM exhibits long-range order exclusively along the replica direction. In contrast, the conventional correlation function computed directly from the RDM remains short-ranged, reflecting the finite correlation length of the underlying SPT wavefunction.

This result has two important implications:
\begin{enumerate}
    \item The bulk strange correlator of a fixed-point SPT wavefunction in \((d+1)\)-dimensions can be obtained by evaluating the twisted Rényi-\(N\) operator on the reduced density matrix, which is a \(d\)-dimensional mixed state. This means that physical observables throughout the bulk can be accessed and simulated via nonlinear operators on the reduced density matrix.

    \item Previous approaches often relied on the entanglement spectrum as a fingerprint for SPT wavefunctions, which is easily obtained in one dimension but numerically challenging in higher dimensions. In contrast, our mapping demonstrates that simpler to compute nonlinear observables derived from the entanglement Hamiltonian can effectively characterize the essential features of these wavefunctions.
\end{enumerate}
\section{Twisted Rényi-N operator for 1d SPT chain}\label{sec:sec_III}

In this section, we analyze the twisted Rényi-\(N\) operator for 1d SPT states. By using the Matrix Product State (MPS) formalism, one can efficiently construct both the 1d SPT ground-state wavefunction and its reduced density matrix. In particular, when a 1d SPT phase is protected by a symmetry group \(G\), its MPS representation encodes a projective representation of \(G\)~\cite{pollmann2012symmetry,chen11a,schuch2011classifying}. Consequently, the reduced density matrix inherits this projective structure, which in turn shapes the structure of the twisted Rényi-\(N\) operator.

Concretely, a 1d MPS (on PBC) may be written as
\[
\bigl|\Psi\bigr\rangle \;=\; \sum_{\{m_i\}} \mathrm{Tr}\left[ A^{m_1} A^{m_2} \cdots A^{m_L} \right] \bigl|m_1, m_2, \ldots, m_L\bigr\rangle,
\]
where each \(A^{m_i}\) is a matrix that depends on the local physical index \(m_i\).
For a MPS invariant under a on-site symmetry group \(G\), a symmetry operation \(g \in G\) on each physical site induces a unitary representation \(U_g\) on the virtual (bond) space. This unitary representation \(U_g\) is generally a \emph{projective representation}, meaning that
\[
U_g U_h \;=\; \omega(g,h)\, U_{gh},
\]
where \(\omega(g,h)\) is a phase factor (2-cocycle) satisfying a consistency condition with respect to the associativity of the group $G$.
When the phase factor  \(\omega(g,h)\) cannot be eliminated by
any redefinition of $U_g$, a certain product of two symmetry operations $g,h$
must differ from the operation corresponding to the product $gh$ by
a phase. This is a hallmark of a non-trivial projective representation,
implying that the MPS belongs to a nontrivial SPT phase.

When examining a subsystem of an infinite chain, tracing out the degrees of freedom in the right half results in the reduced density matrix \(\rho\) for the left half. Because of the finite correlation length, the Hilbert space associated with \(\rho\) effectively reduces to a 0-dimensional mixed state, which is determined by the local configuration near the entanglement cut.
This mixed state naturally inherits the projective representation on the virtual (bond) indices of the MPS. Consequently, any symmetry operation $U_g$ with \(g \in G\) acts projectively on \(\rho\), preserving the same nontrivial cocycle \(\omega(g, h)\).
Within this setup, the RDM \(\rho\) is guaranteed to have degenerate eigenvectors, which are connected by the symmetry charge operator, i.e., charged under the action of $G$.
When constructing the twisted Rényi-\(N\) operator by replicating the density matrix \(N\) times, the largest eigenvector dominates, while the contributions of the other eigenvectors decay exponentially with \(N\). Consequently, in the large-\(N\) limit, only the largest degenerate eigenvectors contribute to the twisted Rényi-\(N\) operator. 
Since these eigenvectors belong to different charge sectors, inserting a charge creation operator \(M_O\) in Eq.~\eqref{eq:twisted} is expected to induce long-range order along the replica direction.  Hence, the projective representation of the MPS plays a critical role in establishing the long-range order of the twisted Rényi-\(N\) operator.

As a preliminary step, we calculate the twisted Rényi-N operator for the 1d Haldane phase, and consider specifically the AKLT state \cite{haldane1983continuum,affleck1988valence}. The fixed-point wavefunction of this phase \cite{chen11a,chen2010local} consists of a tensor product of entangled singlet pairs formed between neighboring sites. This structure leads to a twofold degeneracy in the reduced density matrix across a spatial bipartition reminiscent of a spin-\(\tfrac{1}{2}\) degree of freedom
\begin{align}
    \rho=\frac{1}{2}(|\uparrow\rangle \langle\uparrow|+|\downarrow\rangle \langle\downarrow|),
\end{align}
This density matrix transforms under a projective representation of the \(D_{2}\) symmetry group, generated by \(\pi\)-rotations about the \(S_x\) and \(S_y\) axes. The two fold degeneracy in the eigenvalues of $\rho$ is protected by such projective representation. Applying the dimensional extension mapping described in Sec.~\ref{sec:twisted} yields the 1d wavefunction:

\begin{equation}
|\Psi^{1d}\rangle \;=\; \bigotimes_x \frac{1}{\sqrt{2}}\Bigl(\,\lvert\uparrow\rangle_{R,x}\,\lvert\uparrow\rangle_{L,x+1} \;+\; \lvert\downarrow\rangle_{R,x}\,\lvert\downarrow\rangle_{L,x+1} \Bigr).
\end{equation}

This state resembles the fixed-point wavefunction of the Haldane phase, which can be described as a 1d spin chain with two spin-\(\tfrac{1}{2}\) particles per site. Each unit cell (labeled as $x$) consists of two spins, labeled \(L\) and \(R\), and each link between sites carries a maximally entangled symmetric EPR pair. It is important to note that the wavefunction \(|\Psi^{1d}\rangle\) contains triplet pairs between sites, whereas the standard AKLT wavefunction consists of singlet pairs. This difference arises because the symmetry operators act on the $R$ component of \(|\Psi^{1d}\rangle\) in each unit cell in a conjugate manner (as defined in Eq.~\eqref{eq:strongsymact}), in contrast to the $L$-component, which originates from the ket vector of the density matrix. Consequently, the symmetric wavefunction \(|\Psi^{1d}\rangle\) forms triplet pairs across the bonds. This difference is inconsequential, since one can switch to an alternative basis to reconcile the two descriptions.
Additionally, the weak \(D_{2}\) symmetries of the density matrix \(\rho\) act as \(\pi\)-rotations about the \(\hat{x}\) and \(\hat{y}\) on the $(|\Psi^{1d}\rangle$ state.

As discussed in Refs.~\cite{you2014wave,zhang2022strange}, the state \(|\Psi^{1d}\rangle\) exhibits a nonvanishing strange correlator when spin correlations are measured. Within our mapping, this strange correlator of the fixed-point wavefunction corresponds directly to the twisted Rényi-\(N\) correlation of the density matrix: 
\begin{align}\label{sc}
&C(N)= \frac{\text{Tr}[\rho^{N}\sigma^x\rho^{N}\sigma^x]}{\text{Tr}[\rho^{2N}]}= \frac{\langle \Psi^{\text{trivial}} | \Sigma_x \Sigma_y |\Psi^{1d}\rangle }{\langle \Psi^{\text{trivial}}  |\Psi^{1d}\rangle} \nonumber\\
&\Sigma_x=\frac{\sigma^x_{x,L}+\sigma^x_{x,R}}{2},
\end{align}
with a distance of $|x-y|=N$ between unit cells, and $| \Psi^{\text{trivial}} \rangle=\otimes_x \frac{1}{\sqrt{2}}(|\uparrow\rangle_{L,x} |\uparrow\rangle_{x,R}  + |\downarrow\rangle_{x,L} |\downarrow\rangle_{x,R})$ a tensor product of EPR pairs within each unit cell.
In this construction, the operator $\Sigma_x$ is a symmetrized operator acting on both the \(L\) and \(R\) components within each unit cell.

\subsection{Numerical result for Twisted Rényi-N correlator across the 1d SPT transition}\label{sec:haldane}

To illustrate our theoretical analysis, we investigate the entanglement Hamiltonians of a 1d Haldane chain as it crosses from an SPT phase into a trivial phase. We focus on the half-chain RDM, which behaves like a $0$d ensemble, and compute its twisted Rényi-N correlator. We find that this correlator exhibits long-range order in the replica direction throughout the SPT phase but switches to exponential decay upon entering the trivial phase. Consequently, the twisted Rényi-N correlator of the RDM provides a clear signature for detecting phase transitions.
Importantly, unlike topological measures such as string order parameters that require the full 1d wavefunction, the twisted Rényi‑$N$ correlator depends solely on the RDM, effectively a 0d ensemble. This makes it a more practical and accessible tool for identifying topological transitions.

In particular, we consider the spin-$1$ model
\begin{equation} \label{eq:Haldane_D}
    H=J\sum_j \bm{S}_j\cdot \bm{S}_{j+1} + D\sum_j (S^z_j)^2,
\end{equation}
with $D\geq 0$, whose ground state belongs to either the (gapped) Haldane phase for $D<D_c$, while it becomes topologically trivial for $D>D_c$, with $D_c\approx 0.95J$~\cite{Chen_2003}. The non-trivial topological nature of the former, which lies in the same phase as the AKLT state, is reflected in a double degeneracy of its entanglement spectrum and the existence of spin-\(\tfrac{1}{2}\) edge states. This phase is protected by any of the following symmetries: spatial inversion, time reversal, or the dihedral symmetry group $D_2$ generated by $\pi$ rotations about the \(S_x\) and \(S_y\) axes~\cite{pollmann10,pollmann2012symmetry,schuch2011classifying}. 

We obtain the ground state $|{\psi_D}\rangle$ of $H$ using the infinite density-matrix renormalization group method (iDMRG) \cite{White1992,McCulloch2008}. As mentioned in the previous section, we exploit the finite correlation length $\xi$ within the trivial and Haldane phases to numerically evaluate the twisted Rényi-$N$ correlation $C(N)$. 
Considering the wavefunction on an infinite chain, tracing out the degrees of freedom in the right half results in the reduced density matrix \(\rho\) for the left half.
The finite correlation length \(\xi\) in both phases allows us to approximate the RDM of the half-chain as a 0-dimensional ensemble localized near the entanglement cut. By restricting our analysis to matrix product state (MPS) representations with a finite bond dimension \(\chi\), the RDM can be expressed as:
\begin{equation}
    \rho=\sum_{\alpha=1}^{\chi}\lambda_\alpha |{\alpha}\rangle \langle {\alpha}|.
\end{equation}

In the SPT phase, \( \rho \) exhibits a degenerate spectrum associated with different parity charges under \( R_z(\pi) \). This degeneracy arises from the projective representation structure of the RDM in the SPT wavefunction. Consequently, if we choose an operator that flips the \( R_x(\pi) \) (or \( R_z(\pi) \)) charge in the twisted Rényi-\( N \) correlator, the correlator should remain nonvanishing in the large-\( N \) limit.

\begin{figure}
    \centering   \includegraphics[width=0.8\linewidth]{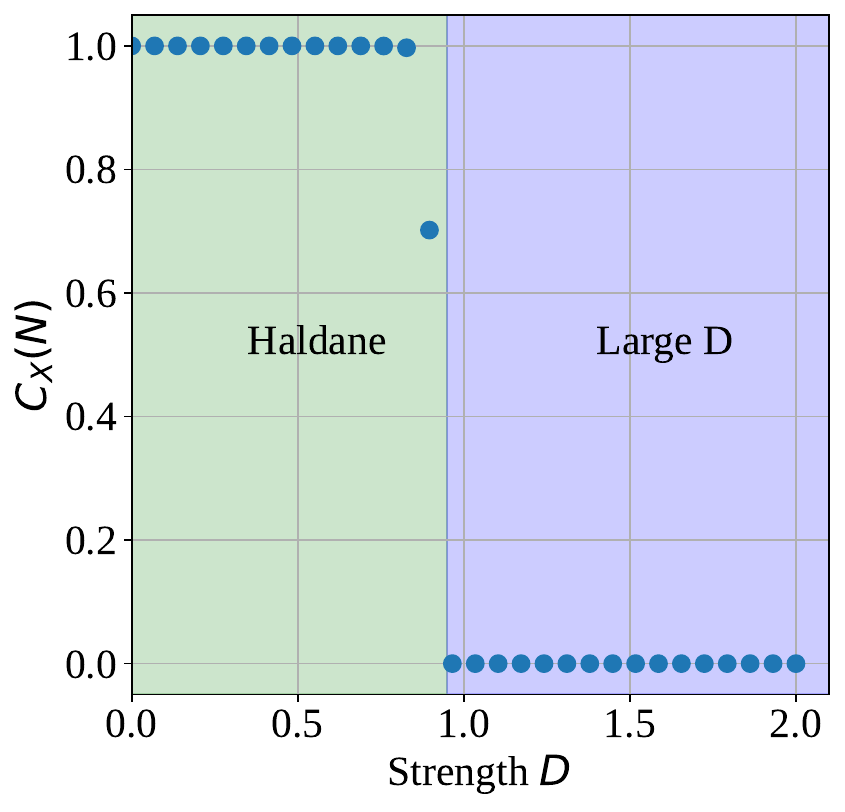}
    \caption{\textbf{Twisted Rényi-$N$ correlation for the Haldane chain.} The ground state is obtained using infinite DMRG with bond dimension $\chi=60$, and $N=100$. }
    \label{fig:Haldane_chain}
\end{figure}

 We evaluate the following twisted Rényi-$N$ correlation:
\begin{equation} \label{eq:CN_gen}
C_X(N)=\frac{\Tr\left[\rho^N \mathcal{O}_X \rho^N \mathcal{O}_X\right]}{\Tr\left[\rho^{2N}\right]},
\end{equation}
where $\mathcal{O}_X$ is the parity-flipping operator acting on a $\chi$-dimensional space and defined via:
\begin{equation}
    \mathcal{O}_X\equiv \sum_{\substack{\alpha :\, q_\alpha=+1\\\beta :\, q_\beta=-1}}(|\alpha\rangle \langle\beta|+ \textrm{H.c.}).
\end{equation}
Numerical results for $C_X(N)$ as a function of $D/J\geq 0$ are shown in Fig.~\ref{fig:Haldane_chain}. First of all, this figure shows that the twisted Rényi-$N$ correlator correctly signals the quantum phase transition, with $C_X(N)$ being finite within the Haldane phase but vanishing within the large-$D$ phase. The quantization of $C_X(N)$ within each phase indicates the degeneracy structure of the entanglement spectrum. In particular, within the Haldane phase, the entanglement spectrum is two-fold degenerate with eigenstates $|\pm\rangle$ corresponding to two different $R_z(\pi)$-parities. Hence, in the limit $N\to \infty$ one finds $\rho^N \approx \lambda^N_0 \left(|+\rangle \langle +|+|-\rangle \langle -|\right)$, with $\lambda_0$ the largest eigenvalue. Using the fact that within this subspace $\mathcal{O}_X=|+\rangle \langle -|+|-\rangle \langle +|$, one then finds 
\begin{equation} \label{eq:CX_N_Haldane}
    C_X(N\to \infty)=\lim_{N\to \infty}\frac{\lambda_0^N\Tr\left(\mathcal{O}_X^2\right)}{ \lambda_0^N\Tr\left(|+\rangle \langle +|+|-\rangle \langle -|\right)} =1.
\end{equation}
Following similar reasoning, one finds that \( C_X(N\to \infty)=0 \) within the large-\(D\) phase. This arises because the entanglement spectrum is non-degenerate in the trivial phase (as it forms a linear representation of \( D_2 \)), so only the eigenvector corresponding to the largest eigenvalue survives in the large-\(N\) limit.
Close to the critical point $D_c$ separating these two phases, we find that a larger $N$ is required to numerically suppress the approximate degeneracy of the largest and second largest eigenvalue.   

Finally, we note that the 1d state \(\Psi^{\text{SPT}}\) obtained via the replica construction in Eq.~(\ref{fixed}) is a fixed-point wavefunction composed of entangled pairs between sites, thus exhibiting zero correlation length along the spatial direction. Consequently, both its strange correlator and the dual twisted Rényi-\(N\) correlator converge to a constant value. 
If we extend this procedure to higher dimensions, for example by replicating the 1d reduced density matrix to form a 2d SPT state, the resulting 2d wavefunction shows zero correlation along the $y$-direction (the replica direction) but retains a finite correlation length along $x$. In such higher‑dimensional cases, the strange correlator can therefore be either power‑law or constant \cite{you2014wave}. We will explore this in more detail in Sec.~\ref{sec:2d}.

\subsection{QFT aspects of Twisted Rényi-N correlator}\label{sec:qft}

In this section, we revisit the duality proposed in Sec.~\ref{sec:twisted} from a field-theory perspective. We show that taking replicas of the reduced density matrix generates a higher-dimensional SPT wavefunction.
As a concrete illustration, we apply this framework to the 1d Haldane chain described in Sec.~\ref{sec:haldane}. However, the formalism is quite general and should apply to most SPT wavefunctions that can be represented as the partition function of a Wess-Zumino-Witten (WZW) theory\cite{xu2013wave,bi2015classification,xu2013nonperturbative}.

The Haldane chain is characterized by an O(3) Nonlinear Sigma Model (NLSM) with a topological term $\Theta = 2\pi $\cite{xu2013wave,xu2013nonperturbative, haldane1983nonlinear}. Its ground state wavefunction can be effectively described using a path integral formulation (see Refs.~\cite{xu2013wave,you2014wave}) via
\begin{align}
|\psi\rangle \sim \int \mathcal{D}[\bm{m}]e^{-S[\bm{m}]}|\bm{m}\rangle 
\label{eq:psi_WZM}   
\end{align}
involving an O(3) rotor $\bm{m}(x)$
\begin{align}
 \mathcal{S}&=\int_{S^1} dx \left(\frac{1}{g}(\partial_x \bm{m})^2+\text{WZW}_1[\bm{m}]\right)
 \label{eq:NLSM}   
 \end{align}
  coupled with a WZW term
 \begin{align}
 \text{WZW}_1[\bm{m}]\equiv\frac{2\pi i}{\Omega^3} \int_0^1 du~\epsilon_{ijk}  m^{i}\partial_x m^{j} \partial_u m^{k}, 
 \end{align}
 with  \(\Omega^3=8\pi\), and where $\bm{m}$ is extended such that $\bm{m}(u=0,x)=(0,0,1)$ and $\bm{m}(u=1,x)=\bm{m}(x)$ to compute this topological contribution.

\begin{figure}[h]
    \centering
\begin{tikzpicture}
\node at (0,0){\includegraphics[scale=0.9]{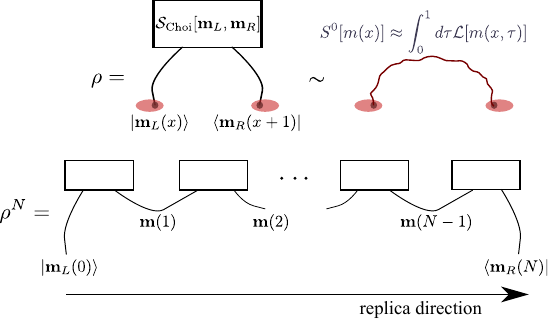}};
\node at (-3.5,2.2) {(a)};
\node at (-4.,-0.) {(b)};
\end{tikzpicture}    
    \caption{\textbf{Twisted Rényi-$N$ correlator from QFT.} Viewing $\rho$ as the path integral of a spin‑$1/2$ field on an interval $[x,x+1]$, its ket and bra vectors serve as boundary fields $|\bm{m}_L(x)\rangle$ and $ \langle\bm{m}_R(x+1)|$. Replicating the density matrix, $\rho \to \rho^{N}$, extends this path integral along the replica ($x$) direction to the interval $[0,N]$.}
    \label{fig:WZW}
\end{figure}

The three components $(m^1, m^2, m^3)$ in the O$(3)$ rotor can be viewed as the spin polarizations in the $S^x$, $S^y$, and $S^z$ directions. 
By applying a spatial bipartition at $x=0$ to the wavefunction in Eq.~\eqref{eq:psi_WZM} and tracing out the region with $x>0$, we obtain (see Fig.~\ref{fig:WZW}a)
\begin{align} \label{eq:rho_WZW}
\rho \sim \int \mathcal{D}[\bm{m}_L, \bm{m}_R]e^{-\mathcal{S}_{\text{Choi}}[\bm{m}_L, \bm{m}_R]}|\bm{m}_L\rangle \langle \bm{m}_R|
\end{align}
with
\begin{align}\label{choidouble}
&\mathcal{S}_{\text{Choi}}=\mathcal{S}[\bm{m}_L]+\mathcal{S^*}[\bm{m}_R]-U \bm{m}_L\cdot \bm{m}_R
\end{align}
We phenomenologically introduce the coupling \( U \) between the ket and bra vectors near the spatial partition cut, which arises from tracing out a subsystem. Since the original wavefunction in Eq.~\eqref{eq:psi_WZM} has short-range correlations, the reduced density matrix effectively behaves like that of a 0-dimensional system. Thus, we only consider the \(\bm{m}_{L/R}\) configurations near the cut.  
The action \(\mathcal{S}[\bm{m}_{L/R}]\) is given by the 0d WZW term, which is known as the Berry phase for spin coherent state: 
\begin{align}
&\mathcal{S}[\bm{m}_a]=\frac{i}{2}\int^1_0 du ~(1-\cos(\theta_a))~\partial_u \phi_a,\nonumber\\
&(m_1,m_2,m_3)=(\sin(\theta)\cos(\phi), \sin(\theta)\sin(\phi),\cos(\theta))
 \end{align}
which characterizes the path integral for spin-$1/2$ particle~\cite{altland2010condensed}.

When the density matrix is replicated into a 1d wave function as in Eq.~\eqref{fixed}, the process closely parallels the coupled-wire construction. In this framework, each unit cell at site $\bm{x_i}$ contains two coupled 0d theories: the left (L) and right (R) components, originating respectively from the bra vector of the $i$-th density matrix and the ket vector of the $(i+1)$-th density matrix.
Moreover, as detailed in Sec.\ref{sec:recover}, the bra and ket spaces of each $\rho$ constitute an entangling block that couples the right component of the $i$-th site with the left component of the $(i+1)$-th site. While the wavefunction in a lattice model is defined only on discrete sites, in the continuum limit, it can be viewed as a continuously varying vector field $m(x)$ that smoothly interpolates across space.
In this formulation, $\rho$ takes the same form as in Eq.~\eqref{eq:rho_WZW}, with the action approximated as:
\begin{align}\label{eq:coupled}
&\mathcal{S}_{\text{Choi}}[\bm{m}(x)]
\\
&\sim \int^{1}_0 dx \left(\frac{1}{g}(\partial_x\bm{m})^2 \right.  \left.+\frac{2\pi i}{\Omega^3} \int_0^1 du~\epsilon^{ijk}  m_{i}\partial_x m_{j} \partial_u m_{k} \right) \nonumber
\end{align}
Here, $\mathcal{S}_{\text{Choi}}$ is interpreted as the path integral of a spin-$1/2$ field defined over the interval $[0,1]$ as in Fig.~\ref{fig:WZW}(a). 
Within this representation, the ket and bra vectors $|\bm{m}_L\rangle$ and $\langle \bm{m}_R|$ serve as boundary fields fixed at the segment endpoints, fulfilling $\bm{m}_L = \bm{m}(0)$ and $\bm{m}_R = \bm{m}(1)$. This aligns with the holographic duality discussed in Sec.~\ref{sec:recover}, wherein the ket and bra vectors correspond to physical degrees of freedom located at adjacent sites.
The index $x$ labels the unit-cell positions introduced by replicating the density matrix along the replica direction. By continuously extending $\bm{m}(x)$ within the interval $[0,1]$, the boundary conditions $\bm{m}_L = \bm{m}(0)$ and $\bm{m}_R = \bm{m}(1)$ simply anchor the field at the boundaries of the region. By exploiting the invariance under global rotation, we phenomenologically extend $U\bm{m}(0)\cdot \bm{m}(1)$ as $(\partial_x \bm{m}(x))^2$ locally into the region $[0,1]$ \cite{altland2010condensed,xu2013wave}, leading to the first line in Eq.~\eqref{eq:coupled}, namely a $1$d WZW model on the finite interval $[0,1]$, with $1/g \sim U$.

When we replicate the density matrix as $\rho \to \rho^N$, the path integral in Eq.~\eqref{eq:coupled} extends along the replica direction (the $x$-direction) over the interval $[0, N]$. This arises as the inner product of two density matrices requires the ket of the first to match the bra of the second, effectively stitching the path integrals from different segments into a continuous 1d path integral. Consequently, $\rho^N$ corresponds to the 1d WZW theory depicted in Fig.~\ref{fig:WZW}(b):
\begin{align}\label{eq:1dcoupled}
    &\rho^N = \int \mathcal{D}[\bm{m}(x)]~e^{-\mathcal{S}_{1d}[\bm{m}(x)]}|\bm{m}_L\rangle \langle \bm{m}_R|,\nonumber\\
    &\bm{m}_L = \bm{m}(0), \bm{m}_R = \bm{m}(N)
\end{align}
$\mathcal{S}_{1d}$ is the 1d WZW theory whose boundaries at $x=0$ and $x=N$ maps to the bra ($\bm{m}_L$) and ket ($\bm{m}_R$) vectors of the replicated density matrix $\rho^N$:  
\begin{align}\label{eq:1dfin0}
& \mathcal{S}_{1d}=\int^{N}_0 dx~\left(\frac{1}{g}(\partial_x\bm{m})^2 \right.  \left.+\frac{2\pi i}{\Omega^3} \int_0^1 du~\epsilon^{ijk}  m_{i}\partial_x m_{j} \partial_u m_{k} \right)
\end{align}
Thus, $\rho^N$ resembles the $O(3)_1$ WZW theory in $1+0$ dimensions, which describes the time evolution of a free spin-$\tfrac{1}{2}$ particle. Since the spin–spin correlations in this WZW theory are long-ranged \cite{you2014wave}, the effective $1+0$d action $\mathcal{S}_{1d}$ likewise exhibits long-range order. Specifically, the correlation function $\langle \mathbf{m}(x_1)\mathbf{m}(x_2)\rangle$ approaches a non-zero constant as $|x_1 - x_2| \to \infty$.

The twisted Rényi-$N$ correlation precisely takes the form of such correlation function
\begin{align}\label{eq:1dfin}
\frac{\int \mathcal{D}[\bm{m}(x)] \bm{m}(x_1) \bm{m}(x_2) e^{-\mathcal{S}_{1d}[\bm{m}(x)]}}{\int \mathcal{D}[\bm{m}(x)]  e^{-\mathcal{S}_{1d}[\bm{m}(x)]}} \sim \mathcal{O}(1) .
\end{align}

As a result, the twisted Rényi-$N$ correlator converges to a finite value, a universal feature for all 1d SPT states whose reduced density matrices effectively behave as 0d mixed states. In this scenario, the coupling term \( U \) (corresponding to \(1/g\) in the WZW theory), which entangles the ket and bra vectors of the RDM in Eq.~\eqref{choidouble}, is always relevant and can thus be taken to $U \rightarrow \infty, g\rightarrow 0$ without loss of generality.

In contrast, for 2d SPT phases, the entanglement Hamiltonian behaves like a 1d mixed state, and the replicated density matrix \(\rho^{N}\) maps onto a 2d nonlinear sigma model (NLSM) accompanied by a WZW term. Under renormalization, the coupling constant \( g \) flows to a finite fixed point~\cite{xu2013nonperturbative}.  
This description aligns with the WZW formulation of strange correlators in 2d SPT phases~\cite{xu2013nonperturbative}, where the system is characterized by a (2+0)-dimensional WZW theory. Here, the spatial gradient term \(\frac{1}{g}(\partial_{i}\mathbf{n})^{2}\) renormalizes to a finite value, resulting in algebraically decaying correlations. We will further explore this scenario in Sec.~\ref{sec:ladder}.

\subsection{Differentiate distinct SPT phases using the twisted Rényi-\(N\) operator}

In our previous discussion, we showed that the twisted Rényi-\(N\) correlator of the entanglement Hamiltonian can effectively distinguish an SPT phase from a trivial phase. Here, we investigate whether this operator can also differentiate between \textit{distinct} SPT phases. To address this question, we consider a spin-2 chain that supports multiple SPT phases under a \((\mathbb{Z}_2 \times \mathbb{Z}_2)^2\) symmetry.  Specifically, we consider a variant of the spin-2 SO(5) AKLT chain introduced in Ref.~\cite{tu2011intermediate}, whose Hamiltonian is given by:
\begin{equation}
H=\sum_{j}\sum_{\gamma =1}^{4}J_{\gamma }(\bm{S}_{j}\cdot \bm{S}%
_{j+1})^{\gamma }+D\sum_{j}(S_{j}^{z})^{2},  \label{eq:Hamiltonian}
\end{equation}
The coupling constants \(J_\gamma\) are precisely fine tuned with \(J_1 = -\tfrac{11}{6}\), \(J_2 = -\tfrac{31}{180}\), \(J_3 = \tfrac{11}{90}\), and \(J_4 = \tfrac{1}{60}\), with \(D \geq 0\). In the special case where \(D = 0\), the model simplifies to the SO(5) topological spin chain investigated in Refs.~\cite{tu2008class,scalapino1998so}. Here, each site carries a spin-2 representation, which arises from the fusion of two spin-\(\tfrac{3}{2}\) degrees of freedom. Each spin-\(\tfrac{3}{2}\) forms an SO(5) singlet by entangling with either its left or right neighbor across the lattice sites.
When \(D = 0\), the Hamiltonian in Eq.~(\ref{eq:Hamiltonian}) reduces to
\[
H_{D=0} = 2\sum_{j} \bigl[P_{2}(j,j+1) + P_{4}(j,j+1)\bigr],
\]
where \(P_{S_T}(j,j+1)\) projects the neighboring sites \(j\) and \(j+1\) onto the total spin-\(S_T\) subspace. This model exhibits SO(5) symmetry and possesses an exact matrix product state (MPS) ground state, as demonstrated in Refs.~\cite{tu2008class,scalapino1998so}. 
To identify the SO$(5)$ symmetry, we begin by considering the \(S^z\) basis \(\{|m\rangle\}\) with \(m = \pm 2, \pm 1, 0\). Let us denote by $L^{ab}$, with $1\leq a<b \leq 5$, the $10$ generators of the SO$(5)$ Lie algebra, and choose two mutually commuting Cartan generators to be  
\[
L^{12} = |2\rangle\langle 2| \;-\; |-2\rangle\langle -2| \quad \text{and} \quad
L^{34} = |1\rangle\langle 1| \;-\; |-1\rangle\langle -1|.
\]  
In addition, we introduce the operators  
\[
L^{15} = \frac{1}{\sqrt{2}}\bigl(|2\rangle\langle 0| + |0\rangle\langle -2| + \mathrm{h.c.}\bigr)
\]    
\[
L^{35} = \frac{1}{\sqrt{2}}\bigl(|1\rangle\langle 0| + |0\rangle\langle -1| + \mathrm{h.c.}\bigr),
\]  
Together with the remaining generators, these satisfy the SO(5) commutation relations,  
\[
[L^{ab}, L^{cd}] \;=\; i\left(\delta_{ac}\,L^{bd} + \delta_{bd}\,L^{ac} \;-\; \delta_{ad}\,L^{bc} \;-\; \delta_{bc}\,L^{ad}\right),
\]  
which fully determine the ten generators \(L^{ab}\) of the SO(5) algebra. Hence, the global SO$(5)$ symmetry of the system at $D=0$ is generated by $L^{ab}_{\textrm{tot}}= \sum_{j} L_{j}^{ab}$.
For \(D > 0\), the SO(5) symmetry is broken, and the Hamiltonian retains only a set of \(Z_2\) symmetries. These arise from its invariance under global \(\mathbb{Z}_2\) rotations generated by  
\[
U^{ab} = e^{i\pi\,L^{ab}}, \quad \text{for all } L^{ab}.
\]  
These \(\mathbb{Z}_2\) symmetries collectively form the group \((\mathbb{Z}_2 \times \mathbb{Z}_2)^2\). Without loss of generality, the generators of this group can be chosen as  
\[
\{I, U^{12}\} \;\times\; \{I, U^{15}\} \;\times\; \{I, U^{34}\} \;\times\; \{I, U^{35}\}.
\]  
This reduced symmetry structure reflects the breaking of the continuous SO$(5)$ symmetry to a discrete subgroup in the presence of a nonzero \(D\).

The phase diagram of this model was previously investigated using the iDMRG method in Ref.~\cite{tu2011intermediate}. As \(D\) increases from \(0\) to \(\infty\), the system undergoes two quantum phase transitions at critical points \(D_{c_1}\) and \(D_{c_2}\). The first transition, at \(D_{c_1}\), leads to the so-called `Intermediate Haldane' (IH) phase, which is topologically equivalent to the spin-1 Haldane chain, and the second, at \(D_{c_2}\), brings the system into a trivial phase.

For \(0 \le D < D_{c_1}\), the phase is referred to as the \textit{SO(5) Haldane phase}, whose salient feature are governed by the SO$(5)$-AKLT point at \(D=0\). In this phase, the system realizes a four-dimensional projective representation under the \((\mathbb{Z}_2 \times \mathbb{Z}_2)^2\) symmetry group, and the edge states behave like spin-\(\tfrac{3}{2}\), resulting in a four-fold degeneracy.

In the intermediate range \(D_{c_1} < D < D_{c_2}\), the IH phase emerges. As \(D\) increases, the uniaxial anisotropy in Eq.~(\ref{eq:Hamiltonian}) progressively suppresses the \(\left|\pm 2\right\rangle\) and \(\left|\pm 1\right\rangle\) spin states, with \(\left|\pm 2\right\rangle\) being most strongly affected. Consequently, the \(S^i_z = \pm 2\) components almost disappear from the ground state, leaving \(\left|\pm 1\right\rangle\) and \(\left|0\right\rangle\). This situation mirrors the spin-1 Haldane phase~\cite{okamoto2011distinguish,schollwock1995haldane,oshikawa1992hidden} studied in Sec.~\ref{sec:haldane}, which hosts spin-\(\tfrac{1}{2}\) edge states. In the IH phase, the \((\mathbb{Z}_2 \times \mathbb{Z}_2)\) subgroup generated by \(L^{12}\) and \(L^{15}\) becomes trivial, whereas the complementary \((\mathbb{Z}_2 \times \mathbb{Z}_2)\) subgroup generated by \(L^{34}\) and \(L^{35}\) retains a projective symmetry characterized by a two-dimensional representation.

\begin{figure}
    \centering    \includegraphics[width=0.85\linewidth]{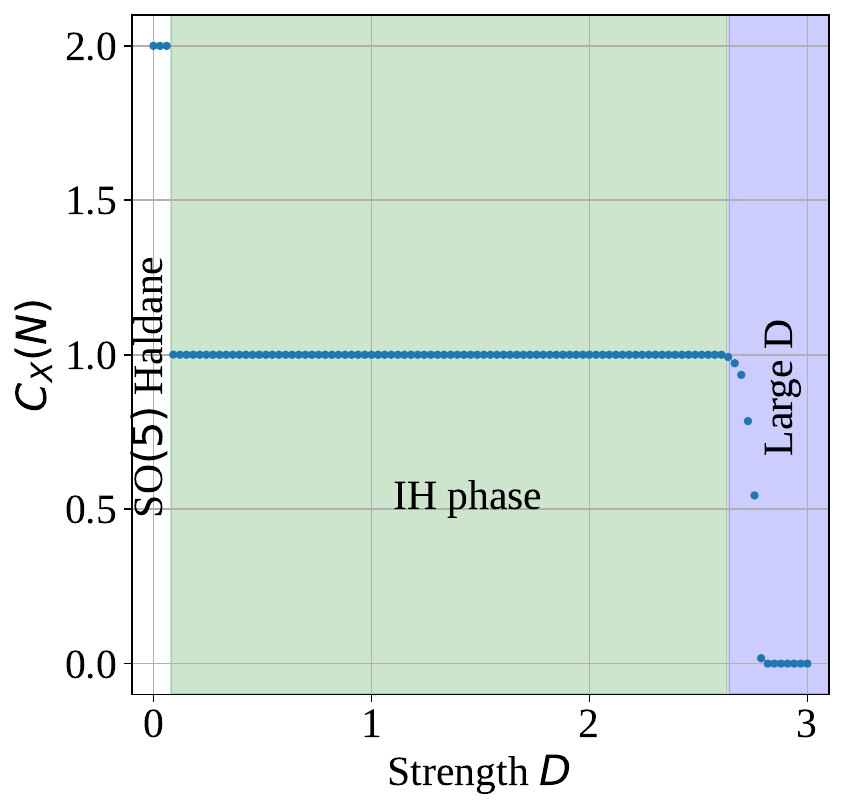}
    \caption{\textbf{Spin-$2$ SO$(5)$ AKLT chain.} Phase boundaries drawn as in Ref.~\cite{tu2011intermediate} with bond-dimension $\chi=60$. This Rényi-$N$ correlator measures the degeneracy of the entanglement spectrum in the limit $N\to \infty$. The system size $L$ has to be taken $L\to \infty$ (numerically large) to suppress the almost degenerate spectrum close to critical points. Here we use $N=100$. }
    \label{fig:spin2_SO5_AKLT}
\end{figure}

Here we evaluate the generalized twisted Rényi-$N$  correlator $C_X(N)$ as defined in Eq.~\eqref{eq:CN_gen}, by inserting the $R_z$-parity flip operator $\mathcal{O}_X$. Numerical results for bond dimension $\chi=50$ and $N=100$ are shown in Fig.~\ref{fig:spin2_SO5_AKLT}. As expected, the twisted Rényi-$N$ correlator is quantized and is able to distinguish different SPT phases. In particular, a similar analysis to the one in the previous version but applied to the SO$(5)$ phase gives
\begin{equation}
    \rho^N\approx \lambda_0^N \sum_{\sigma=\pm 1;m=1,2}|\sigma, m\rangle \langle \sigma, m|
\end{equation}
where $\sigma$ is the parity $R_z$, and $m$ labels the additional symmetry charge that allows to distinguish different eigenstates. Nonetheless, we find that $\mathcal{O}_X=\sum_{\sigma=\pm 1;m, m'=1,2}|\sigma, m\rangle \langle -\sigma, m'|$ correctly captures the different degeneracy of the phases. In particular, $C_X(N\to \infty)=2$ within the SO$(5)$ phase, and $C_X(N\to \infty)=1$ within the IH phase. Finally, $C_X(N\to \infty)=0$ in the trivial phase.

\section{Twisted R\'enyi-N correlator in higher dimensional  systems} \label{sec:sec_IV}

 \subsection{Entanglement Hamiltonian for 2d chiral states}\label{sec:2d}

We now extend the formalism of the twisted Rényi-\(N\) correlator to the reduced density matrix of 2d symmetry-protected topological (SPT) states. In 1d SPT wavefunctions, the manifestation of long-range order in the replica direction is relatively straightforward because the reduced density matrix effectively behaves as a 0d system. In this case, the long-range order is guaranteed by the projective representation of the corresponding 1d matrix product state (since it leads to a degenerate spectrum of the 0d RDM). However, generalizing this framework to higher dimensions introduces a more intricate challenge.
In this section, we focus on the reduced density matrix of a 2d Chern insulator, initially studied in Ref.~\cite{chen2013quantum,qi2012general,lundgren2013entanglement}. The entanglement Hamiltonian of this system indeed resembles that of a 1d chiral fermion. We demonstrate that the twisted Rényi-N operator associated with this reduced density matrix exhibits quasi-long range order along the replica direction. This result provides a new fingerprint for identifying and characterizing topological states in two dimensions through their entanglement properties.

To derive the reduced density matrix of a Chern insulator (or an integer quantum Hall system), we follow the method described in Ref.~\cite{chen2013quantum,qi2012general,lundgren2013entanglement}. This approach utilizes a `cut-and-glue' procedure, where counter-propagating chiral edge modes from two boundaries of the Chern insulator are coupled and integrated into the bulk wavefunction. This framework aligns conceptually with the coupled-wire construction\cite{Kane2014} of a Chern insulator, in which the bulk wavefunction emerges from the coupling of chiral boundary modes.
The procedure begins by dividing the Chern insulator into two disconnected regions, \( A \) and \( \bar{A} \), which are initially independent. Each region hosts chiral edge states, as illustrated in Fig.~\ref{cutandglue}b (e.g., at the bottom of region \( A \) and the top of region \( \bar{A} \)). The critical step involves reintroducing interactions between these regions by gluing them along their shared edges. This is accomplished by activating an interaction Hamiltonian that couples \( A \) and \( \bar{A} \). To simplify the analysis, we consider only interactions between the gapless edge modes, disregarding contributions from the gapped bulk degrees of freedom.

Originally proposed in Ref.~\cite{chen2013quantum,qi2012general,lundgren2013entanglement}, the cut-and-glue method reveals that the entanglement properties of a Chern insulator (C=1) can be effectively described using a quasi-one-dimensional system of coupled chiral modes organized in a ladder-like structure. Building on this framework, Ref.~\cite{lundgren2013entanglement,chen2013quantum} explicitly derives the entanglement Hamiltonian of the Chern insulator in terms of a chiral boson field. Using bosonization techniques, the 1d chiral boson fields are defined as follows: 
\begin{align}
\label{eq:bose1}
\varphi_{L}
=\varphi_{L,0}+n_{L}\frac{2\pi x}{l_x}
+\sum_{k>0}\sqrt{\frac{2\pi}{l_x|k|}}\left(a_{k}e^{ikx}+a_{k}^\dagger e^{-ikx}\right)
\end{align}
Note that \(\bigl[\varphi_{L,0}, n_{L}\bigr] = i\). The variables \(\varphi_{L,0}\) and \(n_{L}\) correspond to the zero-energy (zero-mode) component of the chiral boson field, while \(a_{k}\) describes the oscillatory (finite-energy) modes. As demonstrated in Ref.~\cite{lundgren2013entanglement}, these two components contribute independently to the entanglement Hamiltonian ($H_e$) of a Chern insulator. Consequently, the reduced density matrix ($\rho$) can be expressed as a tensor product of the contributions from the zero-mode part and the oscillatory part:
\begin{align}\label{1drd}
\rho=e^{-H_e}=e^{-H_{\text{zero}}} \otimes e^{-H_{\text{osi}}}.
\end{align}
Using Peschel’s method~\cite{peschel2009reduced,peschel2003calculation}, one can obtain the oscillatory part by effectively treating the system as a one-dimensional harmonic oscillator restricted to the positive-velocity branch:
\begin{equation}\label{eq:He_osc_ch}
 H_{\text{osi}} = \sum_{k>0} w_k a_k^{\dagger}a_k, ~~w_k \approx \frac{4K}{m} |k|.
\end{equation}
This yields the free-Bose distribution for the reduced density matrix:
\begin{align}\label{eq:akdak_ansatz}
 &\mathrm{Tr}\left( a_k^\dagger a_k \rho_\mathrm{osc} \right) = \frac1{e^{w_k}-1}.
\end{align}
Here, \(K,m\) are parameters determined by the microscopic details of the 2d Chern insulator discussed in Ref.~\cite{lundgren2013entanglement}.
The zero-mode part of the entanglement Hamiltonian is:
\begin{equation}
 \rho_\mathrm{zero} = \sum_{N_R} \frac{1}{z} e^{-\frac{4\pi K}{l_x m} n_R^2} |n_R \rangle  \langle n_R|,
\end{equation}
where \(l_x\) denotes the length of the 1d chain parallel to the cut and \(z\) is the appropriate normalization factor. Combining both the zero-mode and the oscillatory parts gives the full entanglement Hamiltonian:
\begin{equation}\label{eq:Hent_ch}
 H_e =\frac{4K}{m} \left[ \frac{\pi }{l_x} n_R^2 + \sum_{k>0} k \left(a_k^{\dagger}a_k \right) \right].
\end{equation}
Accordingly, the reduced density matrix \(\rho=e^{-H_e}\) can be viewed as the thermal state of a one‑dimensional chiral boson at inverse temperature \(\beta=1\).

Because the 1d reduced density matrix of a 2d Chern insulator maps onto this thermal chiral‑boson state, the twisted Rényi‑\(N\) operator naturally appears as the Euclidean‑time two‑point correlator of the chiral boson on a \(\mathsf T^{2}\) torus.  Specifically, we express the replica density matrix \(\rho^{\,N}\) through a path‑integral representation of the 1d chiral boson in imaginary time:
\begin{align}
 \rho^N=\int \mathcal{D}\phi_L ~ e^{-\int^{N}_{0} d\tau \int^{l_x}_{0}  dx~ (\partial_x \phi_L\partial_{\tau} \phi_L+K(\partial_x\phi_L)^2)} 
\end{align}
The path-integral representation can be understood as analogous to the partition function of a 1d chiral boson on a ring of circumference \( l_x \) at an inverse temperature \( \beta = N \). 
Importantly, as with any thermal state, it is crucial to first take the thermodynamic limit \( l_x \to \infty \) before varying the effective temperature \( \beta = N \). This order of limits ensures the system is analyzed correctly within the thermodynamic limit. If instead \( l_x \) remains finite, finite-size effects introduce a gap in the spectrum of the reduced density matrix \(\rho\), resulting in a dominant eigenvector. Subsequently, in the large-\( N \) limit, this dominant eigenvector dictates the behavior of the twisted Rényi-\( N \) correlator, leading to exponential decay in the \( N \)-direction. In our subsequent analysis, we consistently consider the regime \( N \ll l_x \) to avoid these finite-size artifacts.

\begin{figure}[h]
    \centering
\includegraphics[width=0.3\textwidth]{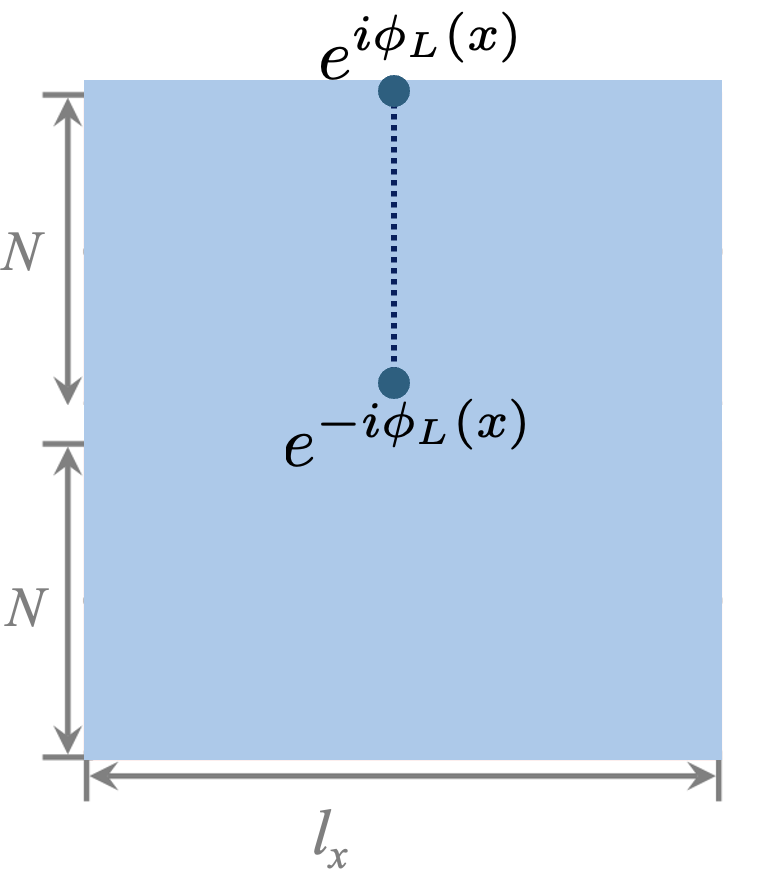}
    \caption{Euclidean path‑integral view: \(\rho^{2N}\) is a chiral‑boson partition function on a \(T^{2}\) torus \((\Delta x=l_x,\;\Delta\tau=2N)\); inserting \(e^{i\phi_L}\) at \((\tau=0,x_0)\) and \((\tau=N,x_0)\) produces the twisted Rényi‑\(N\) correlator, akin to the temporal two‑point function at imaginary‑time separation \(N\).} 
    \label{pathin}
\end{figure}

From this viewpoint, \(\rho^{\,2N}\) corresponds to the Euclidean path integral of a chiral boson on a \(T^{2}\) torus with spatial extent \( \Delta x = l_x \) and imaginary‑time extent \( \Delta \tau = 2N\).  Within this setup, the twisted Rényi‑\(N\) correlator is obtained by inserting the operators \(e^{i\phi_L(0,x_0)}\) and \(e^{i\phi_L(N,x_0)}\) at the same spatial point \(x_0\) but at different imaginary times \(\tau=0\) and \(\tau=N\), respectively.  Hence, it represents the temporal two‑point function of a chiral boson separated by an imaginary‑time interval \(N\).
\begin{align}
   &C(N) =\frac{\text{Tr}[\rho^N e^{i\phi_L(x_0)}\rho^N e^{-i\phi_L(x_0)}]} {\text{Tr}[\rho^{2N}]}\nonumber\\
    &=\langle e^{i\phi_L(0,x_0)} e^{i\phi_L(N,x_0)} \rangle.
    \label{eq:Chiralboson_TRC}
\end{align}
 After first taking the spatial length \(l_x\) to the thermodynamic limit, this `temporal  correlation' decays algebraically with the replica index \(N\). For a free chiral boson, the temporal correlator scales as \(C(N)\sim N^{-1}\), in agreement with the chiral‑CFT results of Refs.~\cite{verlinde1987chiral,sarma2025effective}.
Consequently, we expect the twisted Rényi‑\(N\) correlator to exhibit a power‑law decay along the replica direction, characteristic of quasi‑long‑range order.

\subsection{Entanglement Hamiltonian for 2d SPTs}

The preceding argument also applies to non‑chiral SPT phases, exemplified by the quantum spin Hall (QSH) state whose entanglement Hamiltonian $H_e$ was studied in Ref.\cite{lundgren2013entanglement,chen2013quantum}. In the QSH phase, $H_e$ is equivalent to a 1d helical boson: left‑ and right‑moving modes carry different $U^s(1)$ charges, which forbids backscattering. The entanglement Hamiltonian is invariant under $U^e(1)\times U^s(1)$, with their mixed 't Hooft anomaly enforces a gapless spectrum.

We first consider the entanglement Hamiltonian of a two-dimensional \( U^e(1) \times U^s(1) \) SPT phase, analogous to a quantum spin Hall insulator. Following a derivation similar to Sec.~\ref{sec:2d} and Ref.~\cite{chen2013quantum}, the reduced density matrix of the QSH insulator maps onto that of a one-dimensional Luttinger liquid. Its entanglement Hamiltonian can be expressed in terms of two conjugate bosonic fields \( \phi_1 \) and \( \phi_2 \), satisfying: 
\begin{align}
\left [\phi_1(x),\,\frac{\partial_{x'}\phi_2(x')}{2\pi}\right] = i\delta (x-x')
\label{commutation}
\end{align}
Physically, $e^{i\phi_1(x)}$ creates a boson at point $x$, while the vertex operator $e^{i\phi_2}$ inserts a phase slip. These boson fields act under $U^e(1) \times U^s(1)$ symmetry as:
\begin{align}
U^e(1): \,& \phi_1 \rightarrow \phi_1+\alpha\\ \nonumber
&\phi_2\rightarrow \phi_2\\ \nonumber
U^s(1): \,& \phi_1 \rightarrow \phi_1\\ \nonumber
& \phi_2 \rightarrow \phi_2+\beta
\end{align}
Terms such as \( \cos \phi_{1,2} \), which could potentially gap the Luttinger liquid, are forbidden by these symmetry constraints.

References~\cite{lundgren2013entanglement,chen2013quantum} further demonstrate that the reduced density matrix of the 2d QSH insulator can be viewed as the thermal state of a 1d helical boson. Consequently, the replica density matrix \(\rho^N \) can be formulated as an imaginary-time path integral of a 1d Luttinger liquid
\begin{align}
 \rho^N=\int \mathcal{D}\phi_2 ~ e^{-\int^N_{0} d\tau \int^{l_x}_{0}  dx~ (\frac{K}{2\pi}(\partial_{\mu}\phi_2)^2)} 
\end{align}
Here, \( \phi_2 \) is the vertex operator of the Luttinger liquid, \( K \) is the Luttinger parameter, and within the QSH framework, \( \phi_2 \) is charged under the spin \( U(1)_s \) symmetry.

Thus \(\rho^{\,2N}\) can be viewed as the path integral of a 1d gapless boson on a torus with spatial extent \(\Delta x = l_x\) and imaginary‑time extent \(\Delta\tau = 2N\). In this picture, the twisted Rényi‑\(N\) correlator inserts the \(U(1)_s\)-charged operators \(e^{\pm i\phi_2}\) at \(\tau=0\) and \(\tau=N\) in the path integral, probing the temporal two‑point correlator at an imaginary‑time separation \(\Delta\tau = N\).
As in the case of the chiral boson~Eq.~\eqref{eq:Chiralboson_TRC},
it  exhibits an algebraic decay with exponent determined by the Luttinger parameter \(K\)
when the thermodynamic limit $l_x \to \infty$ is taken first:
\begin{align}\label{eq:twiqsh}
    &C(N)=\frac{\text{Tr}[\rho^N e^{i \phi_{2}(x_0)}\rho^N e^{-i\phi_{2}(x_0)}] }{\text{Tr}[\rho^{2N}]} \nonumber\\
    &=\langle e^{i\phi_2(0,x_0)} e^{i\phi_2(N,x_0)}\rangle=N^{-1/K} .
\end{align}
Notably, similar identical algebraic decay is observed in the strange correlator of the QSH state\cite{lepori2023strange}, in quantitative agreement with our duality framework: the twisted Rényi‑\(N\) correlator of the reduced density matrix faithfully reproduces the bulk strange correlator.
For the free theory we set the Luttinger parameter \(K = 1\); in interacting cases, \(K\) can be tuned by the strength of the interactions.
In Sec.~\ref{sec:ladder}, we will elaborate on this case by numerically studying a coupled‑ladder system using density‑matrix renormalization group (DMRG) calculations.

We emphasize that the algebraic decay of the twisted Rényi-\(N\) correlator in the replica direction holds only if we first take the spatial extent \( l_x \) to infinity before enlarging \( N \). If \( l_x \) remains finite, the reduced density matrix $\rho$ acquires a gapped spectrum (due to finite-size effects) with a dominant eigenvector. In the large-\( N \) limit, this dominant eigenvector controls the twisted Rényi-\( N \) correlator, causing it to exhibit exponential decay along the \( N \)-direction. Specifically, for finite \( l_x \), the asymptotic behavior of the twisted Rényi-\( N \) correlator in Eq.~\eqref{eq:twiqsh} is dominated by \( e^{-N/l_x} \). This aligns with the physical interpretation: for fixed finite \( l_x \), the replica density matrix describes a quantum spin Hall (QSH) state in the thin-stripe (quasi-1d) limit. In 1d no symmetry-protected topological (SPT) phase exists under the \( U^e(1)\times U^s(1) \) symmetry, ensuring any corresponding strange correlator should be short-ranged.

Finally, we discuss the limitations of our theoretical derivation for the twisted Rényi-\( N \) correlator. Our analysis relies on a simplified model of a non-interacting Chern insulator with zero correlation length, where the entanglement Hamiltonian \( H_e \) in Eq.~\eqref{1drd} reduces to a relativistic chiral boson in the infrared (IR) limit.  
However, this approximation breaks down in more general settings. For a general SPT wavefunction with finite correlation length and interactions, the entanglement Hamiltonian becomes more complex than the form given in Eq.~\eqref{1drd}. 
In such cases, the twisted Rényi-\( N \) correlator still exhibits quasi-long-range order, decaying polynomially as \(\sim N^{-\eta}\) along the replica direction. Crucially, the exponent \(\eta\) is non-universal, depending on microscopic details such as interactions and lattice-scale effects. This behavior parallels that of strange correlators in SPT wavefunctions\cite{lepori2023strange}, which also display power-law decay (\(\sim r^{-\eta}\)) with a non-universal exponent. 

\subsection{Numerical simulation from ladder systems} \label{sec:ladder}

Our argument in Sec.~\ref{sec:2d} can be naturally extended to other 2d non-chiral SPT states whose entanglement Hamiltonian exhibits features reminiscent of a 1d gapless theory with helical modes. 
To investigate the entanglement properties of these systems, we employ the cut-and-glue approach introduced in Ref.~\cite{lundgren2013entanglement}. This method demonstrates that for any gapped state, the entanglement is localized at the cut, causing the entanglement Hamiltonian \( H_e \) to resemble a 1d Hamiltonian. Building on this insight, we analyze the entanglement Hamiltonian by placing the system on a ladder geometry and perform a bipartition between the two gapless chains with anomalous symmetry, which are coupled via a rung term. 
In this setup, the ladder system as a whole forms a gapped state with a finite correlation length \cite{poilblanc2010entanglement}. Meanwhile, the entanglement Hamiltonian retains the essential characteristics of a 1d gapless theory, including 't Hooft anomalies.

To validate our prediction regarding the long-range order for the twisted Rényi-$N$ correlator elucidated in the previous section, we consider the following spin-$1/2$ Hamiltonian on a ladder
\begin{equation}
\begin{aligned} \label{eq:H_ladder}
    H = \sum_{\alpha=u,d} \sum_{j=1}^L &\left(S^x_{j,\alpha} S^x_{j+1,\alpha} +S^y_{j,\alpha} S^y_{j+1,\alpha}\right) \\
    &+ J\sum_{j=1}^L \bm{S}_{j,d}\cdot \bm{S}_{j,u},
\end{aligned}
\end{equation}
with $J>0$, and evaluate 
\begin{equation} \label{eq:C_jN}
    C(j,N)=\frac{\textrm{tr}\left(\rho_u^N S^x_{j,u} \rho_u^N S^x_{j,u}\right)}{\textrm{tr}\left(\rho_u^{2N}\right)}.
\end{equation}
Here, $\rho_u$ (or equivalently $\rho_d$) is the reduced density matrix on the upper (lower) chain of the ground state $|{\psi_{\textrm{g.s.}}}\rangle$ of $H$ in Eq.~\eqref{eq:H_ladder}.

The spin Hamiltonian on a ladder can be viewed as two coupled spin-\(\frac{1}{2}\) XY models, interconnected by Heisenberg interactions along the rungs. In particular, the first line of Eq.~\eqref{eq:H_ladder} describes the intra-wire component, which follows the spin-\(\frac{1}{2}\) XY model.  
The entanglement Hamiltonian (\(H_e = -\ln(\rho_u)\)) of this ladder system under rung bipartition was investigated in Ref.~\cite{poilblanc2010entanglement,lundgren2013entanglement,chen2013quantum}. It was found to closely resemble the spin-\(\frac{1}{2}\) XY Hamiltonian, which governs the interactions within each individual chain when the rung coupling is absent. The spin-\(\frac{1}{2}\) XY Hamiltonian supports a free boson with a gapless mode characterized by a central charge \(c = 1\).  

A notable feature of the entanglement Hamiltonian is the presence of a \(U^{V}(1) \otimes U^{A}(1)\) anomaly \cite{chatterjee2025quantized}. To further elucidate this, one can define the conserved charges \(Q^{V}\) and \(Q^{A}\), which generate the \(U^{V}(1)\) and \(U^{A}(1)\) symmetries, respectively:
\begin{align}
&Q^{V}= \sum_i (S^z_i+\frac{1}{2}), \nonumber\\
&Q^{A} = \sum_i \frac{1}{4}(S_{2i-1}^xS_{2i}^y -S_{2i}^y S_{2i+1}^x),
\end{align}
Here \(Q^{V}\) denotes the total \(S^z\) quantum number along the chain. The definition of \(Q^{A}\) is a bit unconventional and it should be regarded as an \textit{emanant} symmetry which flows to the
winding number of the XY model in the continuum~\cite{miao2021conjectures,vernier2019onsager}.
It defines a $U^{A}(1)$ symmetry that is quantized with integer eigenvalues and commutes with the XY model. However, \( U^{A}(1) \) is not an onsite symmetry because it cannot be expressed as a sum of onsite charges. The mixed anomaly between \( U^{V}(1) \) and \( U^{A}(1) \) can be explicitly realized in lattice models, such as a free fermion model or a spin-1/2 XY model, which serves as the entanglement Hamiltonian under consideration. 
As demonstrated in Ref.~\cite{chatterjee2025quantized}, the \( U^{V}(1) \otimes U^{A}(1) \) symmetry accurately captures both the axial anomaly of a 1d free fermion and the ’t Hooft anomaly of a 1d free boson. At low energies, the \( U^{A}(1) \) symmetry effectively acts as the axial \( U(1) \) charge, which measures the difference in charge between the left- and right-moving sectors. Consequently, the entanglement Hamiltonian \( H_e \), derived from the ladder system described in Eq.~(\ref{eq:H_ladder}), can be interpreted as the reduced density matrix of a 2d SPT wavefunction under the symmetry group \( U^{V}(1) \otimes U^{A}(1) \) with mixed anomaly.

We obtain this reduced density matrix by first calculating $|{\psi_{\textrm{g.s.}}}\rangle$ in MPS form using iDMRG with a bond dimension $\chi=60$. We run simulations on a chain of system size $2L$ locating the upper (lower) chain on even (odd) sites. Second, we obtain the reduced density matrix $\rho_u$  on even sites, and compress the resulting matrix product operator (MPO) to bond dimension $\chi_{\textrm{max}}$. We then obtain $\rho_u^N$, by iteratively multiplying $\rho_u$ with itself and compressing the resulting MPO to bond dimension $\chi_{\textrm{max}}$ after each multiplication. Next, we apply the local operator $S^x_j$ to $\rho_u^N$. Finally, we multiply $S^x\rho_u^N$ with itself and similarly for $\rho_u^N$, compressing the resulting MPO to maximum $\chi_{\textrm{max}}$ in both cases. Finally, we take the trace to obtain $C(j,N)$ in Eq.~\eqref{eq:C_jN}. 
Moreover, for sufficiently large system sizes $L$, computing $\rho_N^u$ reaches machine precision zero, which we address by renormalizing the tensors at each step.

\begin{figure}
    \centering
    \begin{tikzpicture}
    \node at (0,0){\includegraphics[scale=0.36]{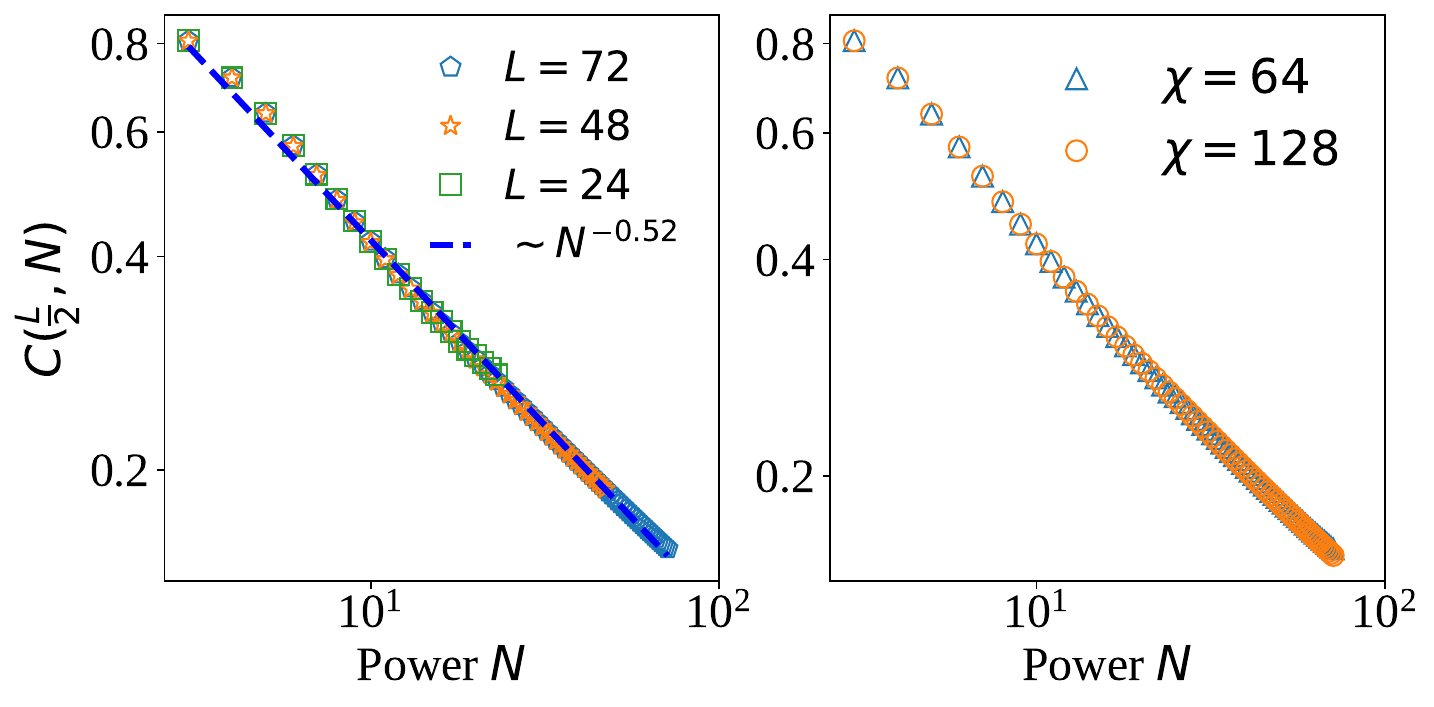}};
    \node at (-3,-1.1) {(a)};
    \node at (1.15,-1.1) {(b)};
    \end{tikzpicture}
    \caption{\textbf{Twisted Rényi-$N$ correlation for ladder.} Numerical evaluation of the twisted Rényi-$N$ correlator in Eq.~\eqref{eq:C_jN} for the ground state of the ladder Hamiltonian in Eq.~\eqref{eq:H_ladder} with $J=6$. The ground state is obtained with DMRG using a bond dimension $\chi=60$. (a) $C(\frac{L}{2},N)$ as a function of $N\leq L$ for various system sizes $L$. In this plot we fix $\chi_{\textrm{max}}=128$, and find a powerlaw decay $C(\frac{L}{2},N)\sim N^{-0.52}$. (b) Convergence in bond dimension for largest system size $L=72$. }
    \label{fig:ladder}
\end{figure}

Figure~\ref{fig:ladder} presents numerical simulations of $C(j,N)$ evaluated at $j=\frac{L}{2}$ for various system sizes $L$. Panel (a) shows that $C(j,N)$ as a function of $N$ collapses across different system sizes (with $N < L$) and approximately decays as a power law in $N$. 
This result matches our analytic prediction in Sec.~\ref{sec:2d}: by identifying the twisted Rényi‑\(N\) correlator with the temporal two‑point correlation of a Luttinger liquid, we used the CFT partition function to show that it decays algebraically. This behavior is also consistent with the strange correlator in symmetry-protected topological (SPT) wavefunctions, which also shows quasi-long-range order ($\sim r^{-\eta}$) as discussed in Ref.~\cite{you2014wave,lepori2023strange}. Panel (b) illustrates the convergence of the numerical results with respect to $\chi_{\textrm{max}}$ for the largest system size considered ($L=72$).

\section{Corollary: LSM theorem in and out of equilibrium}\label{sec:corollary}

The scope of the reduced density matrix and its twisted Rényi-\(N\) correlator can be extended to thermal states and open quantum systems, both of which are formed by mixed-state ensembles, whether in or out of equilibrium.
In this section, we generalize the framework of the reduced density matrix and its twisted Rényi-\(N\) correlator to these cases, employing it as an strengthened version of the original Lieb–Schultz–Mattis (LSM) theorem for closed and open systems.  

Quantum anomalies serve as powerful tools for constraining quantum many-body systems, offering insights into low-energy properties from their ultraviolet (UV) structure. The LSM theorem exemplifies such constraints, revealing universal low-energy behavior dictated by UV symmetries.  
Regardless of the microscopic Hamiltonian (UV sector), the combined spatial symmetry \( U \), internal symmetry \( G \), and filling conditions (e.g., spin per unit cell) impose nontrivial constraints on the infrared (IR) properties of many-body spin systems, enforcing a degenerate or gapless energy spectrum.  
The LSM theorem has deep ties to quantum anomalies. As elucidated in Refs.~\cite{lu2024lieb,jian2018lieb,else2014classifying}, under renormalization group flow, the spatial symmetry \( U \) maps to an effective internal symmetry \( \tilde{U} \) in a coarse-grained, enlarged unit cell. The resulting effective theory-constrained by the LSM theorem and inherently gapless (degenerate)---can be viewed as a quantum field theory with a quantum anomaly under \( \tilde{U} \times G \) symmetry, which again prevents a unique gapped ground state.  
This unified framework connects the LSM theorem to SPT boundary physics and informs the search for quantum spin liquids and deconfined quantum critical points. 

We briefly highlight two corollaries derived from our analysis of the twisted Renyi-N correlator, which can be regarded as a strengthened version of the LSM theorem.

\subsection*{Corollary A: Extension of LSM theorem and the thermal density matrix}

Consider a quantum spin-$1/2$ chain with Hamiltonian $H_\text{1d}$ with short-range interactions. The LSM theorem \cite{lieb1961two,affleck1986proof,oshikawa2000commensurability} states that if $H_\text{1d}$ has both $\text{SO}(3)$ spin-rotation symmetry and lattice translation symmetry, then its energy spectrum must be either gapless or degenerate. A generalized version of the LSM theorem relaxes the $\text{SO}(3)$ symmetry requirement to its subgroups, such as $D_{2h}$.
While the LSM theorem has been rigorously proven, an insightful alternative viewpoint (primary explored in Ref.~\cite{lu2024lieb,jian2018lieb,else2014classifying,cho2017anomaly,po2017lattice}) arises by mapping the effective field theory of $H_\text{1d}$ to the boundary of a 2d SPT phase with $\text{SO}(3) \times \mathbb{Z}_2$ symmetry:
\begin{itemize}
    \item The boundary of this SPT phase can be described by a 1d WZW model, which is also the effective field theory for a 1d spin-$1/2$ chain.
    \item Here, the $\mathbb{Z}_2$ symmetry plays a role analogous to lattice translation in the spin chain that shifts between even and odd sites.
    \item Crucially, the anomaly in the WZW theory under $\text{SO}(3) \times \mathbb{Z}_2$ enforces that the boundary spectrum of the 2d SPT cannot be gapped.
\end{itemize}
These observations also imply that the spin-$1/2$ chain must have either a gapless spectrum or degenerate ground states.

 Exploiting the connection to the boundary physics of a 2d SPT, together with the discussion in Sec.~\ref{sec:ladder}, one can always view the thermal density matrix of 1d spin-$1/2$ chain at unit temperature $\rho=e^{-H_{1d}}$ as the reduced density matrix of a 2d SPT wavefunction protected by \( SO(3) \times \mathbb{Z}_2 \) symmetry. Recalling that the twisted Rényi-$N$ correlator of $\rho$ maps to the strange correlator of the 2d SPT---which is (quasi) long range---we can state the following corollary: 

\begin{shaded}
\textit{For translation-invariant systems with half-integer spin per site and spin-rotation symmetry, its thermal density matrix at unit temperature necessarily exhibits (quasi) long-range order in the symmetric twisted Rényi-N correlator}: 
\begin{align}\label{col}
    &C(N)=\frac{\text{Tr}[e^{-N H_{1d}} \mathcal{O}(x)e^{-N H_{1d}} \mathcal{O^{\dagger}}(x)] }{\text{Tr}[e^{-2N  H_{1d}}]} 
\end{align}
\end{shaded}

Indeed, the replica index $N$ can be viewed as an inverse temperature. From this perspective, the thermal density matrix of the 1d theory is expected to exhibit (quasi) long‑range order along the replica (inverse‑temperature) direction. The operator $\mathcal{O}^\dagger(x)$ is charged under either spin rotation or translation symmetry, so it can be chosen as a spin polarization operator or a valence bond solid (VBS) order parameter. The pair $\mathcal{O}^\dagger(x)$ and $\mathcal{O}(x)$ act at the same spatial location $x$ but are separated along the replica direction.

Hence, while the original LSM theorem ties symmetry and filling constraints to the spectrum of the Hamiltonian (thermal density matrix), our corollary sharpens this link by requiring the thermal density matrix to exhibit LRO along the replica direction. Such LRO necessarily signals a gapless or degenerate spectrum, in line with the LSM theorem. Nonetheless, while LSM only rules out a unique, symmetry‑preserving dominant eigenvector of $\rho = e^{-H_{1d}}$, we predict that the twisted Rényi‑$N$ correlator becomes (quasi) LRO. 
Specifically, if the twisted Rényi-$N$ correlator derived from $\rho = e^{-H_{1d}}$ (for a local Hamiltonian $H_{1d}$) exhibits LRO (or quasi-LRO), it necessarily follows that the energy spectrum of $H_{1d}$ must be gapless or degenerate~\footnote{Notice that the opposite implication is not true, as one can always fine-tune a degenerate mode that has nothing to do with symmetry charge}. To understand this, assume for contradiction that $\rho$ possesses a unique dominant eigenvalue associated with an eigenvector $|\alpha_D\rangle$ invariant under spin-rotation and translation symmetries. In the large-$N$ limit, this eigenvector would dominate the correlator, resulting in short-range correlations. This rules out the existence of a unique symmetry-preserving dominant eigenvector, implying that $H_{1d}$ must be either gapless or degenerate.
Consequently, the presence of (quasi) LRO in the twisted Rényi-$N$ correlator imposes a more stringent condition and could be treated as the strengthened version of LSM Theorem. 

\subsection*{Corollary B: Entanglement LSM theorem and the mixed state density matrix}

The LSM theorem has recently been extended to open quantum systems~\cite{kawabata2024lieb,zhou2025reviving,you2024intrinsic} driven by decoherence and dissipation. Reference~\cite{zhou2025reviving} formulates an entanglement version of LSM theorem: for a mixed state $\rho$ of a spin‑$1/2$ chain with short‑range correlations, weak SO$(3)$ spin‑rotation symmetry, and weak lattice‑translation symmetry (as defined in Eq.~\eqref{eq:weak}), the spectrum of $\rho$ must be either gapless or degenerate. 

Based on our previous discussion, we can state the following corollary for mixed states:

\begin{shaded}
\textit{Consider a short-range correlated mixed state density matrix $\rho$ of a spin-1/2 chain, provided it retains weak translation and weak spin-rotation symmetries, must exhibit LRO or quasi-LRO in its twisted Rényi-$N$ correlator.}
\end{shaded}

This corollary naturally follows from the WZW description of the twisted Rényi-$N$ correlator introduced in Sec.~\ref{sec:qft}. A short-range correlated mixed state of a spin-$1/2$ chain can be represented by two strongly coupled 1d $O(4)_1$ WZW theories~\cite{lee2022symmetry}, with the coupling between the two vector fields (originating from ket and bra spaces) constrained by weak spin-rotation and weak translation symmetries. Upon replicating the density matrix from $\rho$ to $\rho^N$, the path integral along the replica direction effectively realizes a 2d WZW theory, analogous to a 2d SPT wavefunction. Within this framework, the strange correlator of the 2d SPT state maps directly onto the twisted Rényi-N correlator, which displays (quasi-)long-range order for operator charged under either spin-rotation or translation symmetry.

Consequently, for a density matrix exhibiting LRO or quasi-LRO in its twisted Rényi-$N$ correlator, it immediately follows that $\rho$ cannot possess a unique dominant eigenvalue. This result is consistent with the spectral constraints established by the entanglement-LSM theorem in Ref.~\cite{zhou2025reviving}. Importantly, while a gapless density matrix spectrum alone does not guarantee LRO in the twisted Rényi-$N$ correlator, the presence of LRO imposes a stricter constraint than the original LSM theorem.

\section{Possible extension for open quantum systems} \label{sec:sec_V}

Recent work has explored SPT phases in mixed states arising from dissipative and decoherent dynamics~\cite{lee2022symmetry,ma2024symmetry,ma2023topological,sarma2025effective,zhang2022strange}. 
In this section, we extend the scope of the twisted Rényi-$N$ correlator to open quantum systems, demonstrating that the topology of mixed-state symmetry-protected topological (mSPT) phases can be detected using twisted Rényi-N correlators of the lower-dimensional surgery operator. 
To establish this result, we adopt the framework introduced in previous sections and analyze the RDM of the Choi-doubled wavefunction \(|\rho\rangle\rangle\) instead. When the density matrix \(\rho\) belongs to an mSPT phase, \(|\rho\rangle\rangle\) corresponds to a pure SPT wavefunction that exhibits short-range correlations and obeys an area law for entanglement. Since RDM of $|\rho\rangle\rangle$ corresponds to a $(d\!-\!1)$-dimensional mixed state with short-range correlations, the strange correlator of $|\rho\rangle\rangle$ can be directly computed from this RDM.

\begin{figure}
    \centering   \includegraphics[width=1\linewidth]{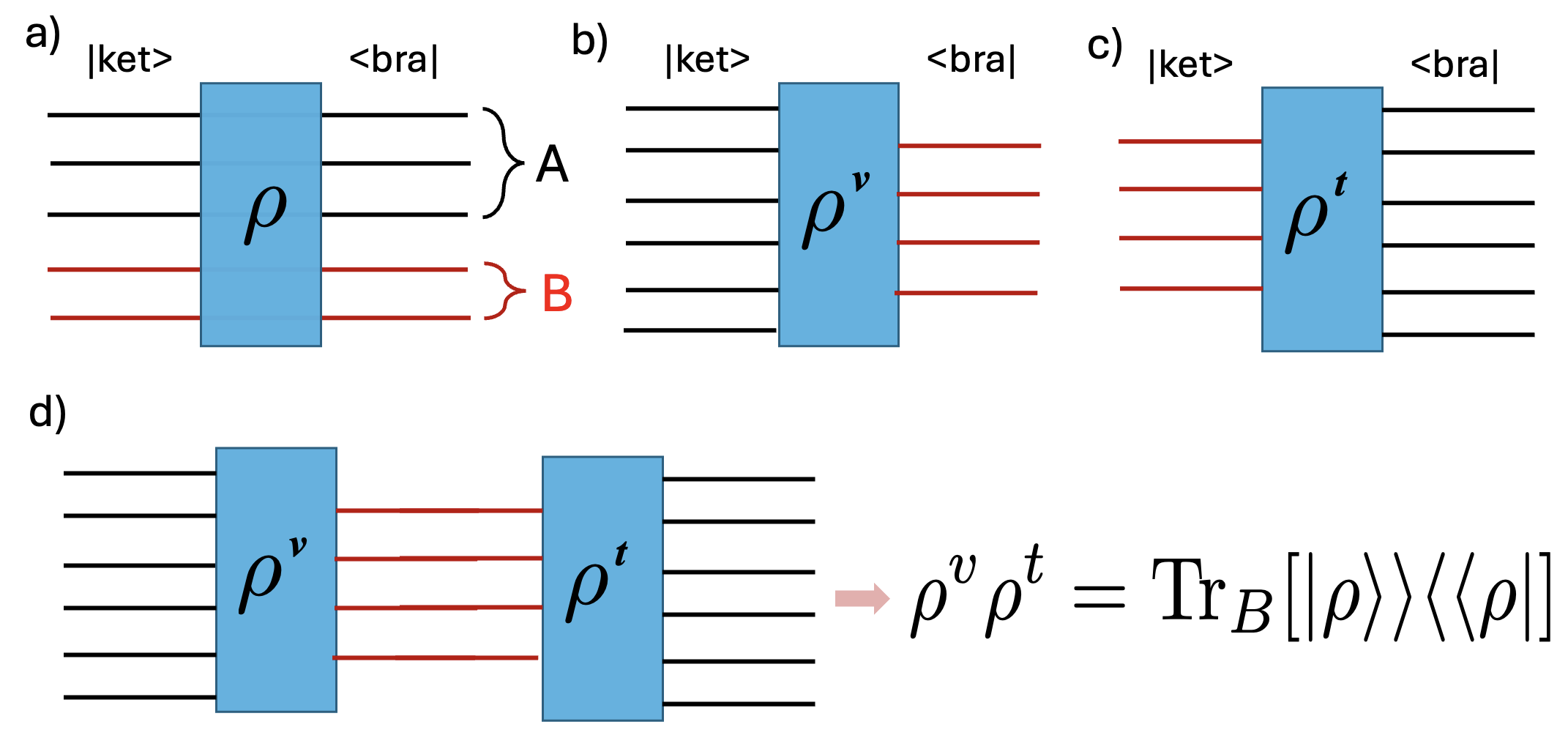}
    \caption{a) The mixed-state density matrix, with the left/right legs representing the ket and bra vectors. Dark/red lines denote the regions \( A \) and \( B \).  
b-c) \( \rho^v \) denotes the vertically surgered density matrix, and \( \rho^t \) denotes the transversely surgered density matrix.  
d) \( \rho^v \rho^t \) is the reduced density matrix obtained from the Choi-doubled state \( |\rho\rangle\rangle \) after tracing out all degrees of freedom in region \( B \). }
    \label{fig:open1}
\end{figure}

We proceed by considering the mixed state SPT described by $\rho$, and perform a bipartition between $A$ and $B$ degrees of freedom such that  
\begin{equation}
    \rho=\sum_{\alpha \beta \alpha' \beta'}\lambda^{\alpha \beta}_{\alpha' \beta'} |A_{\alpha}\rangle |B_{\beta}\rangle \langle A_{\alpha'}| \langle B_{\beta'}|.
\end{equation}
Then, the associated vectorized state in the double Hilbert space is given by
\begin{equation}
    |\rho\rangle \rangle =\sum_{\alpha \beta \alpha' \beta'}\lambda^{\alpha \beta}_{\alpha' \beta'} |A_{\alpha}\rangle_1 |B_{\beta}\rangle_1  |A_{\alpha'}\rangle_2 | B_{\beta'}\rangle_2.
\end{equation}
The subscript $|~\rangle_{a=1,2}$ denotes the layer index in the doubled Hilbert space.
From here, the RDM $\rho^{\text{Choi}}_A=\textrm{tr}_B(|\rho\rangle \rangle \langle \langle \rho|)$ of $|\rho\rangle \rangle$ (up to a normalization factor) is given by
\begin{align}
    &\rho^{\text{Choi}}_A = \sum_{\alpha, \tilde{\alpha}, \alpha', \tilde{\alpha}'} C^{\alpha',\tilde{\alpha}'}_{\alpha,\tilde{\alpha}}
    |A_{\alpha}\rangle_1 |A_{\alpha'}\rangle_2 \langle A_{\tilde{\alpha}}|_1   \langle A_{\tilde{\alpha}'}|_2.\nonumber\\
&C^{\alpha',\tilde{\alpha}'}_{\alpha,\tilde{\alpha}}=\sum_{\beta, \beta'}\lambda^{\alpha \beta}_{\alpha' \beta'}(\lambda^{\tilde{\alpha} \beta}_{\tilde{\alpha}' \beta'})^*   
\end{align}

To relate this RDM to the original mixed-state $\rho$, we define two distinct types of \textit{surgery operators}:
\begin{align}
       \rho^v&=\sum_{\alpha \beta \alpha' \beta'}\lambda^{\alpha \beta}_{\alpha' \beta'} |A_{\alpha}\rangle_1 |A_{\alpha'}\rangle_2 \langle B_{\beta}|_1 \langle B_{\beta'}|_2,\\ \nonumber
           \rho^t&=\sum_{\alpha \beta \alpha' \beta'}(\lambda^{\alpha \beta}_{\alpha' \beta'})^* |B_{\beta}\rangle_1 |B_{\beta'}\rangle_2 \langle A_{\alpha}|_1 \langle A_{\alpha'}|_2=(\rho^v)^\dagger
\end{align}
as illustrated in Fig.~\ref{fig:open1}.
 Here $\rho^v$ is a surgery operator illustrated in Fig.~\ref{fig:open1}b and $\rho^t$ denotes its conjugate transpose, where neither $\rho^v$ nor $\rho^t$ are necessarily square matrices. It then follows that the product $\rho^v\rho^t$ in Fig.~\ref{fig:open1}d) corresponds to the reduced density matrix obtained from the Choi-doubled state $|\rho\rangle \rangle$ after tracing out all degrees of freedom in region $B$, i.e., $\rho^{\text{Choi}}_A = \rho^v \rho^t$. Thus, $\rho^v\rho^t$ corresponds to a lower dimensional mixed state.

\begin{figure}
    \centering   \includegraphics[width=0.8\linewidth]{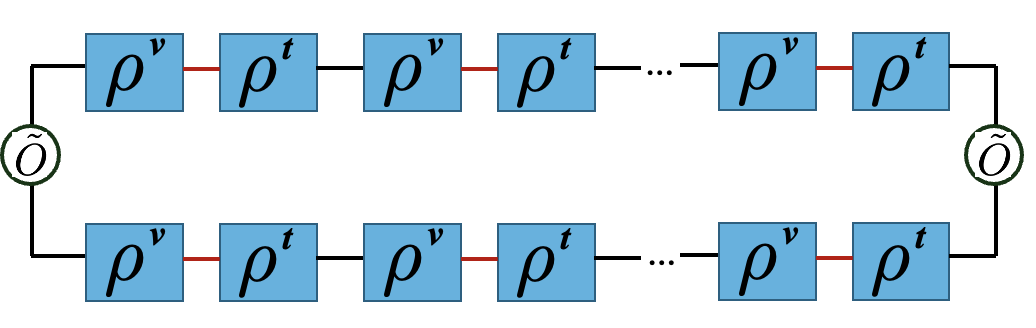}
    \caption{Graphical representation of the twisted Rényi-\(N\) correlator using the surgery operator \(\rho^v \rho^t\). }
    \label{fig:open2}
\end{figure}

Let us for example, consider a 2d system, with site coordinates $(x,y)$. Similarly to Sec.~\ref{sec:Sec_II}, we can then build up a 2d system by gluing 1d RDM $\rho^{\text{Choi}}_{1d}$, such that the bra component of the \(i\)-th density matrix $\rho^{\text{Choi}}_{1d}$, and the ket component of the \((i+1)\)-th define together a unit cell (labeled by the index $x=i+\frac{1}{2}$). The only difference is that now each RDM $\rho^{\text{Choi}}_{1d,x+{1/2}}$ contributes with a doubled Hilbert space on both unit cells $x$ and $x+1$. Namely, denoting by $\mathcal{H}$ the on-site Hilbert space of the original system, then $\rho^{\text{Choi}}_{1d}$ is a density matrix acting on an on-site Hilbert space $\mathcal{H}^{\otimes 2}$. As a result, we can formulate the twisted Rényi-\(N\) correlator using the surgery operator \(\rho^{\text{Choi}}_{1d}=\rho^v \rho^t\) as in Fig.~\ref{fig:open2}, which reproduces the strange correlator for a fixed point SPT wavefunction $|\phi^{\text{fix}}\rangle \rangle$ defined on Choi-double space:
\begin{equation} \label{eq:Open_twist}
\begin{aligned}
&\frac{\langle \langle \phi^0|O(x_1,y) O^{\dagger}(x_N,y) |\phi^{\text{fix}}\rangle \rangle}{ \langle \langle \phi^0|\phi^{\text{fix}}\rangle \rangle}\\
&= \frac{\Tr[M_O(y)(\rho^v\rho^t)^{N} M_O(y
)(\rho^v\rho^t)^{N}]} {\Tr[(\rho^v\rho^t)^{2N}]},
\end{aligned}
\end{equation}
where
\begin{equation}
\begin{aligned} \nonumber
 &|\phi^{\text{fix}}\rangle \rangle = \\
 &\bigotimes_x 
\sum_{\substack{\alpha_x,\tilde{\alpha}_x\\ \alpha'_x,\tilde{\alpha}'_x}} 
C^{\alpha'_x,\tilde{\alpha}'_x}_{\alpha_x,\tilde{\alpha}_x} 
\underbrace{|A_{\alpha_x}^R\rangle_{1,x-\frac{1}{2}} |A_{\alpha'_x}^R\rangle_{2,x-\frac{1}{2}} 
|A_{\tilde{\alpha}}^L\rangle_{1,x+\frac{1}{2}} |A_{\tilde{\alpha}'}^L\rangle_{2,x+\frac{1}{2}}}_{\text{unit cell $x$}}
\end{aligned}
\end{equation}
and
\begin{equation}\nonumber
 |\phi^{0} \rangle \rangle\sim \bigotimes_x (\sum_{\lambda_x}
|\lambda_x^L \rangle_{1,x} |\lambda_x^R \rangle_{1,x})\otimes (\sum_{\lambda'_x}
|\lambda'^L_x \rangle_{2,x} |\lambda'^R_x \rangle_{2,x}).
\end{equation}
The first line in Eq.~\eqref{eq:Open_twist} evaluates the strange correlator of a pure SPT wavefunction $|\phi^{\text{fix}}\rangle \rangle$ defined in 2d Choi double space, where the correlation is measured between operators separated along the spatial $x$ direction. Here, $|\phi^{\text{fix}}\rangle$ denotes a fixed-point state constructed by replicating the 1d density matrix $\rho^{\text{Choi}}_A = \rho^v \rho^t$, following the procedure introduced in Sec.~\ref{sec:twisted}. The superscripts $L/R$ label the left/right component within the same unit cell.
The state $|\phi^0\rangle \rangle$ is a trivial product state in the Choi double space, representing onsite-entangled EPR pairs defined on each unit cell such that the first and second factors are EPR pairs defined on $\mathcal{H}^L_1 \otimes \mathcal{H}^R_1$ and $\mathcal{H}^L_2 \otimes \mathcal{H}^R_2$ respectively. The Hilbert space is doubled as $\mathcal{H}_1 \otimes \mathcal{H}_2$. The operator $O$ can act simultaneously on both layers.
The second line computes the twisted Rényi-$N$ correlator along the replica direction. The operator $\rho^v \rho^t$ defines a 1d mixed state localized near the entanglement cut (along $y$), while $M_O$ acts at a fixed spatial location but is twisted along the replica index $N$.

The twisted Rényi-$N$ correlator, computed from the surgered density matrix, is equivalent to the strange correlator of the SPT state $|\phi^{\text{fix}}\rangle \rangle$. Importantly, $|\phi^{\text{fix}}\rangle \rangle$ is a pure state which lives in the doubled Hilbert space and need not correspond to a valid physical density matrix under the inverse Choi-isomorphism, as it may not be positive semi-definite. Nevertheless, since $|\phi^{\text{fix}}\rangle \rangle$ carries nontrivial SPT order, its strange correlator will exhibit (quasi-)long-range correlations. Therefore, we conjecture that measuring the twisted Rényi-$N$ correlator of $\rho^v \rho^t$ offers a practical diagnostic for detecting and distinguishing mSPT phases.

Finally, we note that Refs.~\cite{sarma2025effective, lee2022symmetry, zhang2022strange} introduced the type-II strange correlator as a diagnostic of SPT order in mixed-state ensembles:
\begin{equation}
\begin{aligned}  \label{eq:type2_strcor}
&\frac{\Tr[O(r) O^{\dagger}(r') \rho O^{\dagger}(r') O(r) \rho^0 ]} {\Tr[\rho \rho^0 ]} \\ 
&= \frac{\langle \langle \rho^0|O_1(r) O_1^{\dagger}(r') \otimes O^T_2(r)O_2^{*}(r') |\rho\rangle\rangle}{\langle \langle \rho^0|\rho\rangle\rangle},  
\end{aligned}  
\end{equation}where \(|\rho\rangle\rangle\) denotes the Choi representation of the density matrix \(\rho\), interpreted as a pure state in a doubled Hilbert space $\mathcal{H}_1\otimes \mathcal{H}_2$ on which the operator $O_1(r)\otimes O_2^T(r)$ acts on the first and second layer. Hence, the Type-II strange correlator corresponds to the strange correlator evaluated on the Choi-double state \(|\rho\rangle\rangle\). 
Comparably, our twisted Rényi-$N$ operator in Eq.~(\ref{eq:Open_twist}) is dual to the strange correlator of the fixed-point wavefunction defined in the Choi-double space, $|\phi^{\text{fix}}\rangle\rangle$. Although this wavefunction does not exactly correspond to the Choi isomorphism image of the original mSPT density matrix, both cases are expected to exhibit qualitatively similar quasi-long-range order in the strange correlator. Therefore, the twisted Rényi-$N$ operator provides an alternative approach for probing mSPT phases.

\section{Conclusions and outlook} \label{sec:Outlook}

In this work, we showed that the twisted Rényi-$N$ correlator, as defined via the reduced density matrix $\rho$ of an area law many-body state $|\Psi\rangle$, serves as a powerful tool for identifying SPT phases. 
In particular, we demonstrated that the Rényi-$N$ correlator of $\rho$ is dual to the strange correlator of the SPT wavefunction $|\Psi\rangle$, establishing a direct connection between the two.

After reviewing the holographic framework introduced in Ref.~\cite{sun2024holographic}, which constructs a $(d+1)$-dimensional fixed‑point SPT state $|\Psi\rangle$ by replicating its $d$-dimensional RDM $\rho$ as $\rho^N$, we numerically and analytically demonstrated the ability of the Rényi-$N$ twisted correlator to identify SPT phases and even to distinguish between different SPT states and detect phase transitions among them in spin chains. We then extended our analysis to 2d systems, including Chern insulators and quantum spin Hall-type states, using both analytical arguments and numerical tensor-network simulations. Moreover, we analytically supported these results by developing a quantum field theory description for $\rho^N$ in the limit of large $N$ as appearing in the twisted Rényi-$N$ correlator by exploiting a mapping to Wess-Zumino-Witten models. This analysis showed that such a correlator is expected to develop either long or quasi-long-range order along the replica index.

In addition, we extended this framework to thermal states and open quantum systems, first deriving an alternative (slightly more general) proof of the Lieb–Schultz–Mattis theorem for both closed and open systems; and finally showing that the topology of mixed-state symmetry-protected topological phases can be characterized by using twisted Rényi-$N$ correlators of the lower-dimensional operator. 

The twisted Rényi‑$N$ correlator provides a powerful framework for simplifying the computation of correlators in a $d$‑dimensional wavefunction by evaluating nonlinear observables derived from its $(d-1)$‑dimensional reduced density matrix. From a holographic perspective, key features of the original wavefunction, such as anomalies and topological invariants, can be inferred directly from the RDM.
Looking ahead, advances in open quantum systems, where decoherence and dissipation drive novel dynamical phases, highlight the importance of nonlinear density matrix functionals. Originally developed to probe replica long-range order, the twisted Rényi‑$N$ operator could be adapted to such systems, potentially revealing new non-equilibrium phenomena.

After completing this work, we noted an independent study that investigates the RDM of SPT wavefunctions through an alternative perspective~\cite{guforth}.

\acknowledgments 
This work was performed in part at the Aspen Center for Physics (YY), which is supported by the National Science Foundation grant PHY-2210452 and the Durand Fund (YY). YY acknowledges support from NSF under award number DMR-2439118. P.S. acknowledges support from the Caltech Institute for Quantum Information and Matter, an NSF Physics Frontiers Center (NSF Grant No.PHY-1733907), and the Walter Burke Institute for Theoretical Physics at Caltech. 
FP acknowledges support by the Deutsche Forschungsgemeinschaft (DFG, German Research Foundation) under Germany’s Excellence Strategy EXC-2111-390814868, TRR 360 (project-id 492547816), and the Munich Quantum Valley, which is supported by the Bavarian state government with funds from the Hightech Agenda Bayern Plus.
The work of MO was partially supported by JSPS KAKENHI Grants No. JP24H00946 and No. JP23K25791, and JST CREST Grant No. JPMJCR19T2.


%

 \end{document}